\documentclass{aastex631}

\usepackage{amsmath}
\usepackage[T1]{fontenc} 
\usepackage{array}
\usepackage{multirow}

\begin{document}

\title{PyMerger: Detecting Binary Black Hole mergers from Einstein Telescope Using Deep Learning}

\author[0000-0002-8266-3005]{Wathela Alhassan}
\affiliation{Particle Astrophysics Science and Technology Centre, Nicolaus Copernicus Astronomical Center,\\ Rektorska 4, 00-614 Warsaw, Poland}

\author[0000-0003-2045-4803]{T. Bulik}
\affiliation{Astronomical Observatory, University of Warsaw, Aleje Ujazdowskie 4, 00-478 Warsaw, Poland\\}


\author{M. Suchenek}
\affiliation{Particle Astrophysics Science and Technology Centre, Nicolaus Copernicus Astronomical Center,\\ Rektorska 4, 00-614 Warsaw, Poland}







\begin{abstract}


We present \textit{PyMerger}, a Python tool for detecting binary black hole (BBH) mergers from the Einstein Telescope (ET), based on a Deep Residual Neural Network model (ResNet). ResNet was trained on data combined from all three proposed sub-detectors of ET (TSDCD) to detect BBH mergers. Five different lower frequency cutoffs ($F_{\text{low}}$): 5 Hz, 10 Hz, 15 Hz, 20 Hz, and 30 Hz, with match-filter Signal-to-Noise Ratio ($MSNR$) ranges: 4-5, 5-6, 6-7, 7-8, and >8, were employed in the data simulation. Compared to previous work that utilized data from single sub-detector data (SSDD), the detection accuracy from TSDCD has shown substantial improvements, increasing from $60\%$, $60.5\%$, $84.5\%$, $94.5\%$ to $78.5\%$, $84\%$, $99.5\%$, $100\%$, and $100\%$ for sources with $MSNR$ of 4-5, 5-6, 6-7, 7-8, and >8, respectively. The ResNet model was evaluated on the first Einstein Telescope mock Data Challenge (ET-MDC1) dataset, where the model demonstrated strong performance in detecting BBH mergers, identifying 5,566 out of 6,578 BBH events, with optimal SNR starting from 1.2, and a minimum and maximum $D_{L}$ of 0.5 Gpc and 148.95 Gpc, respectively. Despite being trained only on BBH mergers without overlapping sources, the model achieved high BBH detection rates. Notably, even though the model was not trained on BNS and BHNS mergers, it successfully detected 11,477 BNS and 323 BHNS mergers in ET-MDC1, with optimal SNR starting from 0.2 and 1, respectively, indicating its potential for broader applicability.

\end{abstract}


\section{Introduction} \label{sec:intro}

\begin{figure}
    \centering
	\includegraphics[scale=0.7]{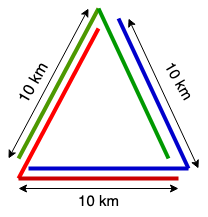}
    \caption{Simplified layout of the three core sub-detectors of the ET telescope.}
    \label{fig:ET}
    \end{figure}

This is a continuation of our previous work \citep{10.1093/mnras/stac3797} (hereafter referred to as WTM1) on the detection of binary black holes (BBHs) gravitational wave (GWs) signals from the Einstein Telescope (ET) using deep learning (DL). ET \citep{Punturo_2010, abernathy2011einstein, 2020JCAP...03..050M} is designed\footnote{See \url{https://www.et-gw.eu/index.php/relevant-et-documents} for relevant documents on the ET design.} to be an underground GWs detector, where the seismic noise is much lower, and hence the level of Newtonian noise. ET as shown in Figure \ref{fig:ET}, will consist of three nested arms (hereafter referred to as sub-detectors) with 10 Km long each, in an equilateral triangle shape, with arm-opening angles of $60^\circ$ \citep{Hild_2010, 2023JCAP...07..068B}. Compared to 3 km length for advanced Virgo \citep{Acernese_2015} and KAGRA \citep{2013PhRvD..88d3007A, 10.1093/ptep/ptaa125}, and 4 km for advanced LIGO \citep{Aasi_2015}, the ET 10 km V-shaped sub-detectors length will significantly improve its sensitivity allowing the observation of GWs at lower frequency that reach $\approx 2$ Hz \citep{Hild_2011}. Each sub-detector of ET will consist of two interferometers, one optimised for low-frequency and one for high-frequency sensitivity. In this work, it is assumed that each sub-detector has a single interferometer.

    The current standard method for detecting GW signals is match-filtering, considered the most sensitive algorithm targeting compact binary sources, such as BBHs with spin-aligned components on quasi-circular orbits \citep{PhysRevD.87.104028, PhysRevD.93.084029, PhysRevLett.115.051101}. One of the main drawbacks of such algorithms is their computational complexity, especially when targeting eccentric BBHs systems \citep{PhysRevD.94.024012}. In addition, match-filtering algorithms might miss signals generated by compact binary populations in dense stellar environments \citep{PhysRevD.90.084016, PhysRevD.93.042004, PhysRevD.95.024038}. These limitations open the door to potential applications of deep learning algorithms, which do not require waveform templates for the detection.
    
    The application examples of deep learning in astrophysics, and more specifically in GWs, were shown in WTM1. Notably, more work has been presented since then, highlighting the significant interest in this promising technology for addressing the considerable challenge posed by the vast amount of data expected from current and upcoming detectors. The Spy project \citep{Glanzer_2023} applied machine learning for glitch classification using data up to the end of the third observing run of LIGO. In this analysis, 233,981 glitches from LIGO Hanford and 379,805 glitches from LIGO Livingston were classified into morphological classes. A related study employed generative adversarial networks (GANs) to simulate hundreds of synthetic images representing the 22 most common types of glitches observed in the LIGO, KAGRA, and Virgo detectors. A neural network model was then used to classify the simulated glitches, achieving an average accuracy of $99.0\%$ \citep{Powell_2023}. Accelerating Artificial Intelligence models for rapid GW detection was recently conducted by \cite{chaturvedi2022inference}. In their work, the ThetaGPU supercomputer was leveraged with the NVIDIA TensorRT-optimized AI ensemble to process a month of advanced LIGO data; the process took 50 seconds. In \cite{Yan_2022}, a novel model called a gravitational wave squeeze-and-excitation shrinkage network (GW-SESNet) was inspired by attention mechanisms, a mechanism in deep learning used to highlight important features, for the classification of real signals, noise, and glitches. GW-SESNet utilized the coherent signal-to-noise ratio (SNR) from multiple detectors and achieved recall values of 0.84, 0.98, and 0.98 for noise, real signals, and glitches, respectively. The role of advanced machine learning tools in the coming decades, either independently or in tandem with traditional data analysis techniques, may become crucial considering the vast amount of the data expected \citep{Shah_2023}.

    The detection of GW sources represents the first step in gravitational wave (GW) data analysis; hence, quick and fast detection is required for subsequent steps, such as source parameter estimation or alerting other telescopes for multi-messenger follow-up and studies of the detected source. Fast detection can also be employed for data marking, facilitating a detailed Bayesian parameter estimation analysis of the relevant part of the detector output at a later stage. Coherent multi-detector observation, in which multiple detectors are coupled together, is beneficial for improving the coherent SNR \citep{macleod2016fully}, making it a current trend in gravitational wave signal detection.
    
In WTM1, our work focused on the detection of BBHs sources that were generated using only one single sub-detector of ET, with low-frequency-cutoff (starting frequency of the binary systems, which increases as the binary system evolves and the components get closer to each other) of 30 Hz, and with five match-filter SNR (hereafter referred to as $MSNR$) ranges: 4-5, 5-6, 6-7, 7-8 and >8. Expanding upon our previous work, we explore the detection efficiency of BBHs using data combined form all the three proposed sub-detectors of ET (hereafter referred to as three sub-detectors coherent data (TSDCD)), with five different lower frequency cutoff ($F_{low}$): 5 Hz, 10 Hz, 15 Hz, 20 Hz and 30 Hz, employing identical $MSNR$ ranges for each frequency. In addition, we used the same luminosity distance ($D_{L}$) range of 145 Mpc - 120 Gpc as in WTM1 for our BBH sources. The main objective is to compare the previous obtained results from a single sub-detector data (hereafter referred to as SSDD), to the results from TSDCD.

The structure of this work goes as follow: in Section \ref{sec:data} we describe the method and tools were used for the generation of TSDCD and the prepossessing pipeline employed for spectrogams preparation. Section \ref{sec:cnn} discusses Deep Residual Neural Networks, and the training and evaluation processes. Evaluating ResNet model on noncontinuous and continuous TSDCD is presented in section \ref{sec:eval} and \ref{sec:eval2}. Evaluation on the first ET mock data challenge is described in section \ref{sec:mdc}. Description of PyMerger software is presented in section \ref{sec:pymerger}. Finally, a summary of this work and a future work plans is presented in Section \ref{sec:con}.

\section{ET data simulation}
\label{sec:data}
The strain in each sub-detector of ET $h_{E\alpha}(t)$ is given as :

\begin{equation}
 h_{E\alpha}(t) =  F_\alpha^+(\theta, \phi, \psi) h_+ (t) + F_\alpha^\times(\theta, \phi, \psi) h_\times(t), 
 \label{eq:1main}
\end{equation}
where $\alpha$ is the index of the sub-detector; $\alpha = (1,2,3)$. $F_\alpha^+$ and $F_\alpha^\times$ are antenna response functions for the two GW polarizations of each sub-detector $\alpha$, which depend on the sky localization (right ascension $\theta$ and declination $\phi$), polarization angle $\psi$, and take into account the motion of the Earth \citep{1998PhRvD..58f3001J}. For a single sub-detector, $F^+$ and $F^\times$ at time $t$ and an angel $\eta$ between the interferometer arms are given as:

\begin{equation}
F_+(t) = \sin \eta [a(t) \cos 2\psi + b(t) \sin 2\psi], \label{eq:1a} 
\end{equation}

\begin{equation}
F_\times(t) = \sin \eta [b(t) \cos 2\psi - a(t) \sin 2\psi], \label{eq:1b}
\end{equation}

where:
\begin{equation}
\begin{split}
a(t) ={}& \frac{1}{16}\sin 2\gamma(3-\cos 2\lambda)(3-\cos 2\delta) \cos[2(\alpha-\phi_r-\Omega_r t)] \\
& - \frac{1}{4}\cos 2\gamma\sin\lambda(3-\cos 2\delta) \sin[2(\alpha-\phi_r-\Omega_r t)] \\
& + \frac{1}{4}\sin 2\gamma\sin 2\lambda\sin 2\delta\cos[\alpha-\phi_r-\Omega_r t] \\
& - \frac{1}{2}\cos 2\gamma\cos\lambda\sin 2\delta\sin[\alpha-\phi_r-\Omega_r t] \\
& + \frac{3}{4}\sin 2\gamma\cos 2\lambda\cos 2\delta,
\end{split}
\label{eq:2a}
\end{equation}

\begin{equation}
\begin{split}
b(t) ={}& \cos 2\gamma\sin\lambda\sin\delta\cos[2(\alpha-\phi_r-\Omega_r t)] \\
& + \frac{1}{4}\sin 2\gamma(3-\cos 2\lambda) \sin\delta\sin[2(\alpha-\phi_r-\Omega_r t)] \\
& + \cos 2\gamma\cos\lambda\cos\delta\cos[\alpha-\phi_r-\Omega_r t] \\
& + \frac{1}{2}\sin 2\gamma\sin 2\lambda\cos\delta\sin[\alpha-\phi_r-\Omega_r t],
\end{split}
\label{eq:2b}
\end{equation}


where $\lambda$ denotes the latitude of ET's location, which we defined same as VIRGO. $\Omega_r$ is Earth's rotational angular velocity, and $\phi_r$ is the phase defining the position of the Earth in its diurnal motion at $t=0$.

$\gamma$ determines the orientation of the detector arms and is measured counter-clockwise from the east to the bisector of the interferometer arms. For the ET in the triangular configuration, $\eta = 60^\circ$.

The polarizations of the GW $h_+ (t)$ and $h_\times(t)$ in Eq \eqref{eq:1main}, from an inspiraling compact binary system such as BBH, are given as: 
\allowdisplaybreaks
\begin{equation}
h_{+}(t) = -\frac{1 + \cos^2 \iota}{2} \left( \frac{G \mathcal{M}}{c^2 D_L} \right) \left( \frac{t_c - t}{G \mathcal{M} / c^3} \right)^{-1/4} 
\cos \left[ 2\Phi_c + 2\Phi(t - t_c; M, \mu) \right] 
\end{equation}
\begin{equation}
h_{\times}(t) = -\cos \iota \left( \frac{G \mathcal{M}}{c^2 D_L} \right) \left( \frac{t_c - t}{G \mathcal{M} / c^3} \right)^{-1/4}
\sin \left[ 2\Phi_c + 2\Phi(t - t_c; M, \mu) \right]
\end{equation}
where $\iota$ denotes inclination angel of the orbital plane of the binary system relative to the observer. $c$ and $G$ represent the speed of light and the  gravitational constant respectively. The quantity $\mu$ denotes the binary system's reduced mass. $t_c$ and $\Phi_c$ represent, respectively, the time and the phase at which the waveform terminates. $t-t_c$ represents the time difference between the current time $t$ and the coalescence time $t_c$. The function $\Phi(t - t_c; M, \mu)$ represents the binary system's orbital phase. $M$ is the source chirp mass, which depend on the mass component of the binary system $m_1$ and $m_2$, and is given as: 

\allowdisplaybreaks
\begin{equation}
  M = (m_1 m_2)^{3/5}/(m_1 + m_2)^{1/5}   
\end{equation}

In WTM1, our emphasis was on simulating BBH sources using a single sub-detector of ET. In this work, we replicated the exact BBH source parameters, data generation tools, and methods employed in WTM1 \citep[for details, please see][section 2]{10.1093/mnras/stac3797} for each sub-detector of ET to obtain TSDCD.

\subsection{Simulation}
We obtained the mass components and redshift ($z$) for simulating BBH systems from \citep{2020A&A...636A.104B}. A broad set of compact binary models was developed by \citet{2020A&A...636A.104B} using the population synthesis code StarTrack \citep{2002ApJ...571..394B, 2002ApJ...572..407B, 2002ApJ...571L.147B}, which has been significantly upgraded in \citet{2008ApJS..174..223B} to focus on processes leading to the formation and further evolution of compact objects.


Restricting the BBH masses to 15-56 \(M_\odot\) and redshift \(z\) values from 0.033 to 11 allows us to focus on low and medium masses within our selected MSNR ranges. In addition, we have considered five \(F_{\text{low}}\) bounds, namely 5 Hz, 10 Hz, 15 Hz, 20 Hz, and 30 Hz. Other parameters such as: the sky position angles ($\theta$ and $\phi$), $\iota$, $\psi$, and coalescence phases, were sampled from a uniform distribution as in WTM1.

 
Assuming the standard $\Lambda$CDM cosmology \citep{2016A&A...594A..13P}, where the Hubble constant $H_0 = 67.74 \, \text{km s}^{-1} \, \text{Mpc}^{-1}$, energy density parameter $\Omega_\Lambda = 0.6910$, and matter density parameter $\Omega_m = 0.3075$. We convert the redshift $z$ to $D_L$ using the analytical approximation provided by \cite{2012PThPh.127..145A} for the $D_L$ between the BBH source and the detector, given by:

\begin{equation}
d_L(z, \Omega_m) = \frac{2c}{H_0 \sqrt{\Omega_m}} (1 + z) 
\left\{ \Phi\left(x(0, \Omega_m)\right) - \frac{1}{\sqrt{1 + z}} \Phi\left(x(z, \Omega_m)\right) \right\},
\end{equation}

where \( c \) is the speed of light in vacuum, and \( \Phi(x) \) is a rational fitting function given by:

\begin{equation}
\Phi(x) = \frac{1 + 1.320x + 0.4415x^2 + 0.02656x^3}{1 + 1.392x + 0.5121x^2 + 0.03944x^3},
\end{equation}

and \( x(z, \Omega_m) \) is defined as:

\begin{equation}
x(z, \Omega_m) = \frac{1 - \Omega_m}{\Omega_m} \cdot \frac{1}{(1 + z)^3}.
\end{equation}

We utilized the frequency-domain phenomenological non spinning BBH model $\text{IMRPhenomD}$ \citep{PhysRevD.93.044006}, which is part of LALSuite \citep{lalsuite}, employing its Python interface \citep{swiglal}. IMRPhenomD is designed for generating GW signals from merging BBH with non-precessing spins. For each sub-detector, using the parameters above, we individually generated a BBH GW signal ($h(t)_{E1}, h(t)_{E2}, h(t)_{E3}$), then added them into the simulated Gaussian detector noise $n$, assuming an additive, Gaussian and stationary noise model. This process results in tow datasets: The data $d_{1,2,3}$ (if the signal $h$ is present) and $d_n$ (if the data contains noise only), can be written as follow:

\allowdisplaybreaks
\begin{align}
    d_{1}(t) &= h_{E1}(t) + n_{1}(t) \\
    d_{2}(t) &= h_{E2}(t) + n_{2}(t) \\
    d_{3}(t) &= h_{E3}(t) + n_{3}(t) \\
    d_{n}(t) &= n_{1}(t) + n_{2}(t) + n_{3}(t)
\end{align}


\begin{figure*}[h!]  
 
    \centering
    \includegraphics[width=0.9\textwidth, height=0.35\textheight]{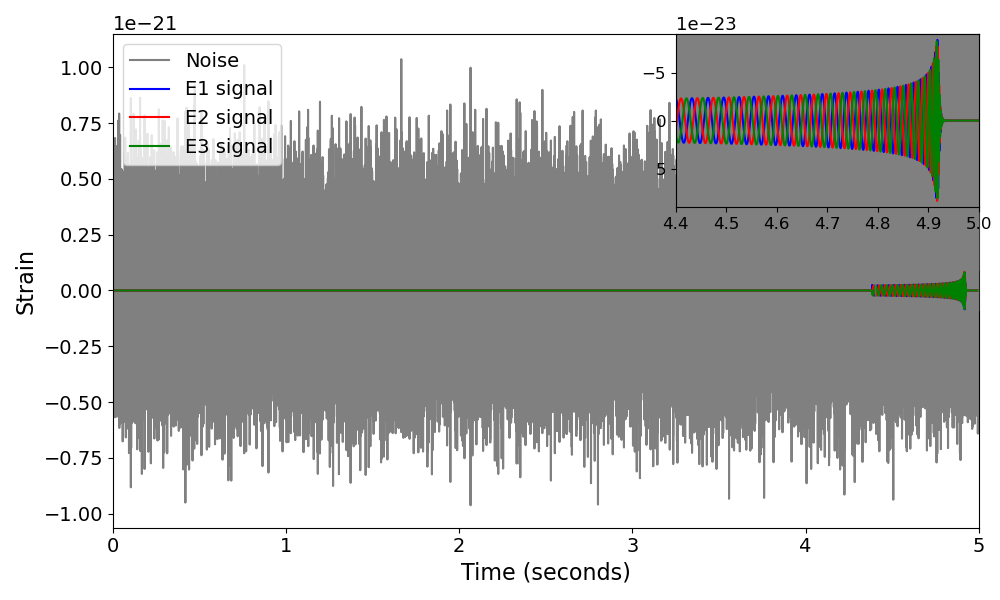}

  \caption{A 5-second segment of simulated noise from a single sub-detector is overlapped with BBH GW signals. The gray line represents a noise from a single sub-detector noise $n_{1}(t)$. The blue, green, and red lines represent the BBH signals from E1, E2, and E3, respectively. The simulated BBH signal has an MSNR in the range $7-8$ with component masses $m_1 = 20 \, M_\odot$ and $m_2 = 25 \, M_\odot$, and $F_{\text{low}}$ of 30 Hz.
}
  \label{fig:ts_3arms_sample}
\end{figure*}

To illustrate how the GW signal would look and compare it to the detector's noise, Figure \ref{fig:ts_3arms_sample} shows the difference in phase between signals from the three sub-detectors and the noise from a single sub-detector in the background. The improvement in SNR through the utilization of data from all sub-detectors of ET combined can be assessed by calculating the coherent SNR. Coherent SNR $\text{SNR}_{\text{coh}}$ enhances the observed signal's SNR by incorporating the coherent combination of multiple detectors \citep{macleod2016fully}. To quantify the improvement in MSNR resulting from the combination of each sub-detector, we calculated $\text{SNR}_{\text{coh}}$ using the following formula:

\begin{equation}
    \text{SNR}_{\text{coh}} = \sqrt{\sum_{\alpha=1}^{3} \text{MSNR}_\alpha^2}
\end{equation}

\begin{figure*}[h!]
  \centering
    \includegraphics[width=0.85\textwidth, height=0.3\textheight]{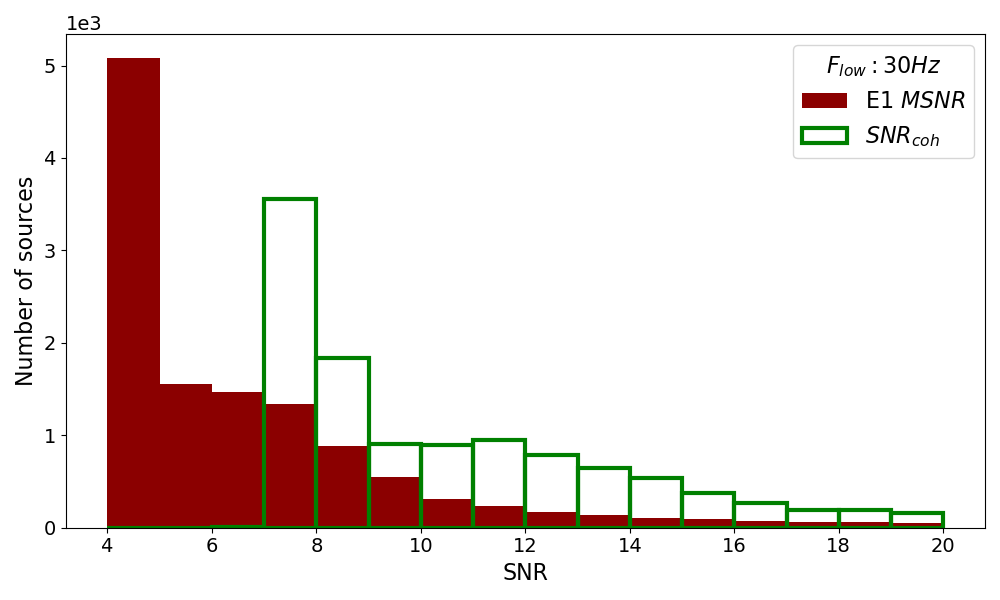}

  \caption{$\text{SNR}_{\text{coh}}$ from the combination of each sub-detector.}
  \label{fig:coh_snr}
\end{figure*}

\subsection{Spectrograms}

\begin{figure*}[h!]
    \centering
	\includegraphics[width=0.67\textwidth, height=0.28\textheight]{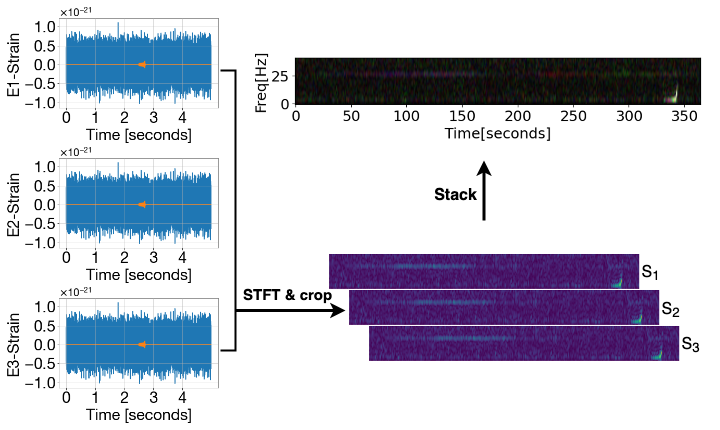}
    \caption{Generation and processing of the spectrograms from TSDCD}
    \label{fig:data_prep}
\end{figure*}

The Short Time Fourier Transform (STFT) \citep{6473982} was used to generate our TSDCD spectrogram samples ($S_{E123}$), as shown in Figure \ref{fig:data_prep}. Three spectrograms ($S_{E1}$, $S_{E2}$, and $S_{E3}$) were produced for each injected segemnets, $d_{1,2,3}(t)$ from each sub-detector ($E1$, $E2$, and $E3$) respectively, and then stacked to form an RGB image $S_{E123}$, represented as:

\begin{equation}
    S_{E123} = 
    \begin{bmatrix}
        R(i, j) \\
        G(i, j) \\
        B(i, j) \\
    \end{bmatrix}
\end{equation}

Where $R(i, j)$, $G(i, j)$ and $B(i, j)$ are the normalized intensity values for each channel at pixel $i, j$ in $S_{E1}(i, j)$, $S_{E2}(i, j)$, and $S_{E3}(i, j)$, calculated as:

\allowdisplaybreaks
\begin{align}
    R(i, j) &= \frac{S_{E1}(i, j)}{\text{max\_intensity}} \\
    G(i, j) &= \frac{S_{E2}(i, j)}{\text{max\_intensity}} \\
    B(i, j) &= \frac{S_{E3}(i, j)}{\text{max\_intensity}}
\end{align}

    
    


where $\text{max\_intensity}$ represents the maximum intensity value across all $S_{E1}$, $S_{E2}$, and $S_{E3}$. A total of 125,000 sources were generated, with each of the five defined $F_{low}$ groups containing 25,000 sources. These sources were evenly distributed among our five $MSNR$ intervals. It is important to note that $MSNR$ values vary between the three detectors using the same parameters due to orientation and polarization. For each source, at least one detector must have the correct $MSNR$ value to be accepted into the $MSNR$ interval.

Similarly, to generate only-noise spectrograms, the following steps were taken: 1) A Gaussian noise with a color matching the power spectrum density of ET, sampled at ET's frequency, was generated for each sub-detector. 2) Three random segments of 5 seconds each were selected from each sub-detector's noise, $n_{1}(t)$, $n_{2}(t)$, and $n_{3}(t)$. 3) STFT was performed on each segment to produce a spectrogram. 4) All three only-noise spectrograms were stacked together.

%

\section{ResNet: Deep Residual Neural Networks}
\label{sec:cnn}
In WTM1, the performance of the most popular five computer vision neural networks models were evaluated for the detection of BBH GW signals, namely VGG-16 and VGG-19 \citep{https://doi.org/10.48550/arxiv.1409.1556}, DenseNet \citep{https://doi.org/10.48550/arxiv.1608.06993}, ResNet \citep{https://doi.org/10.48550/arxiv.1512.03385, wu2018deep}. The ResNet model showed the best overall performance, and thus, our goal in this study is to assess its performance on TSDCD compared to SSDD. 

One of the great advantages of ResNet network, is its ability to overcome the vanishing gradient problem, which considered as one of the challenging issues in training deep and complex neural networks models \citep{2018arXiv180106105H, Roodschild2020ANA}. This was achieved by introducing residual blocks, which enable the training of hundreds of layers \citep{goceri2019analysis, 8936083, Hasan2021}. These blocks use an additive attribute to merge previous and future layers. ResNet is composed of several residual blocks. Each residual block is formed by adding a residual connection every two layers of conventional convolution. The key concept behind the residual connection, considered a pivotal breakthrough in deep learning, is to learn a residual function capable of representing the difference between the input \(X\) and the target \(\mathcal{F}(X)\) (the desired mapping learned by the layer). A residual mapping \(\mathcal{R}(X)\) is learned as follows:
\begin{equation}
\mathcal{R}(X) = \mathcal{F}(X) - X.
\end{equation}

Then, the sum of the residual mapping and the input is the output of the layer (\(Y\)), and can be calculated as:
\begin{equation}   
Y = X + \mathcal{R}(X).
\end{equation}

In the ResNet architecture, \(Y\) is equivalent to \(2\mathcal{F}(X) - X\), reflecting the residual connection \(Y = f(x) + f(x) - x\), where \(f(x)\) represents the output of a series of trainable layers.

Due to this connection, the original output will be superimposed before sending it to the future layer. And hence, mitigate and prevent any gradient explosion or disappearance when using the backpropagation \cite[for details about backpropagation please see]{Munro2010}, and allow for training of very deep networks. More details about ResNet can be found in appendix \ref{sec:resnet}. As previously mentioned in WTM1, ResNet is available in various architectures, distinguished by the number of layers. This includes ResNet-50 and ResNet-101, which comprise 50 and 101 convolutional layers, respectively. We adopt the ResNet-101 architecture (hereafter referred to as ResNet) in this work, as was previously done in WTM1.



\subsection{ResNet training}

We kept the same settings for training the ResNet model as in our previous work (WTM1). However, the dataset size has increased fivefold in TSDCD compared to the single sub-detector data (SSDD) we used before. This expansion is a result of incorporating five different $F_{low}$ values. In machine learning (ML), traditionally, the data is split into three different sets for training, testing, and evaluating the model \citep{10.5555/2380985, raykar2015data, 8463779, nguyen2021influence, 2021arXiv210604525T}. 

Table \ref{tab:no_sample} displays the total number of TSDCD samples utilized for ResNet model training, testing, and validation. All $F_{low}$ values and $MSNR$ ranges are evenly represented in the randomly chosen testing and validation sources, to ensure an unbiased evaluation of the model's performance. The three sub-detector combined spectrograms consist of three layers, where each layer belongs to one sub-detector. Hence, the input layer of ResNet was adjusted to a shape size of 365 × 42 × 3, which describes the width, height, and number of layers of the input image, respectively. A batch size of 256, learning rate of 0.0001, and RMSprop optimizer \citep{babu2020performance, xu2021convergence, elshamy2023improving} were used for training the ResNet model for 200 epochs.

\begin{table}
\centering
\begin{tabular}{ccccc}
\hline
  Type & Number of Sample & Train & Test& Val\\
  \hline
  Injected & 125,000 & 85,000& 20,000 & 20,000  \\ 
  Only noise & 125,000 & 85,000 & 20,000 & 20,000 \\
  \hline
  Total & 250,000 & 170,000& 40,000 & 40,000 \\
  \hline
\end{tabular}
\caption{Total number of TSDCD spectrograms: injected and only-noise spectrograms for training, testing and validation.}
\label{tab:no_sample}
\end{table}

\section{TSDCD evaluation}
\label{sec:eval}

\begin{figure*}[h!]
  \centering

    \includegraphics[width=0.85\textwidth, height=0.27\textheight]{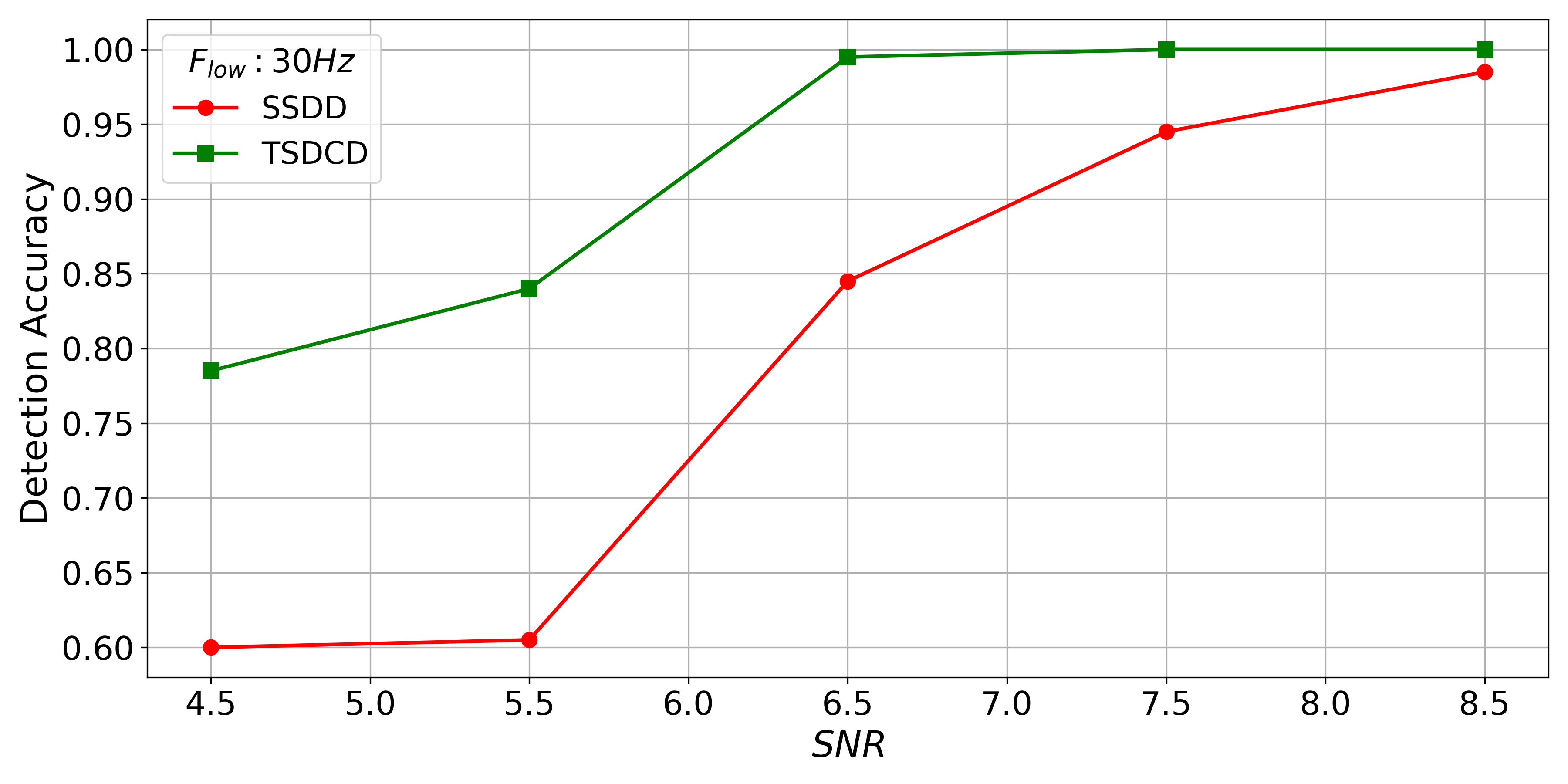}\\
    \includegraphics[width=0.85\textwidth, height=0.27\textheight]{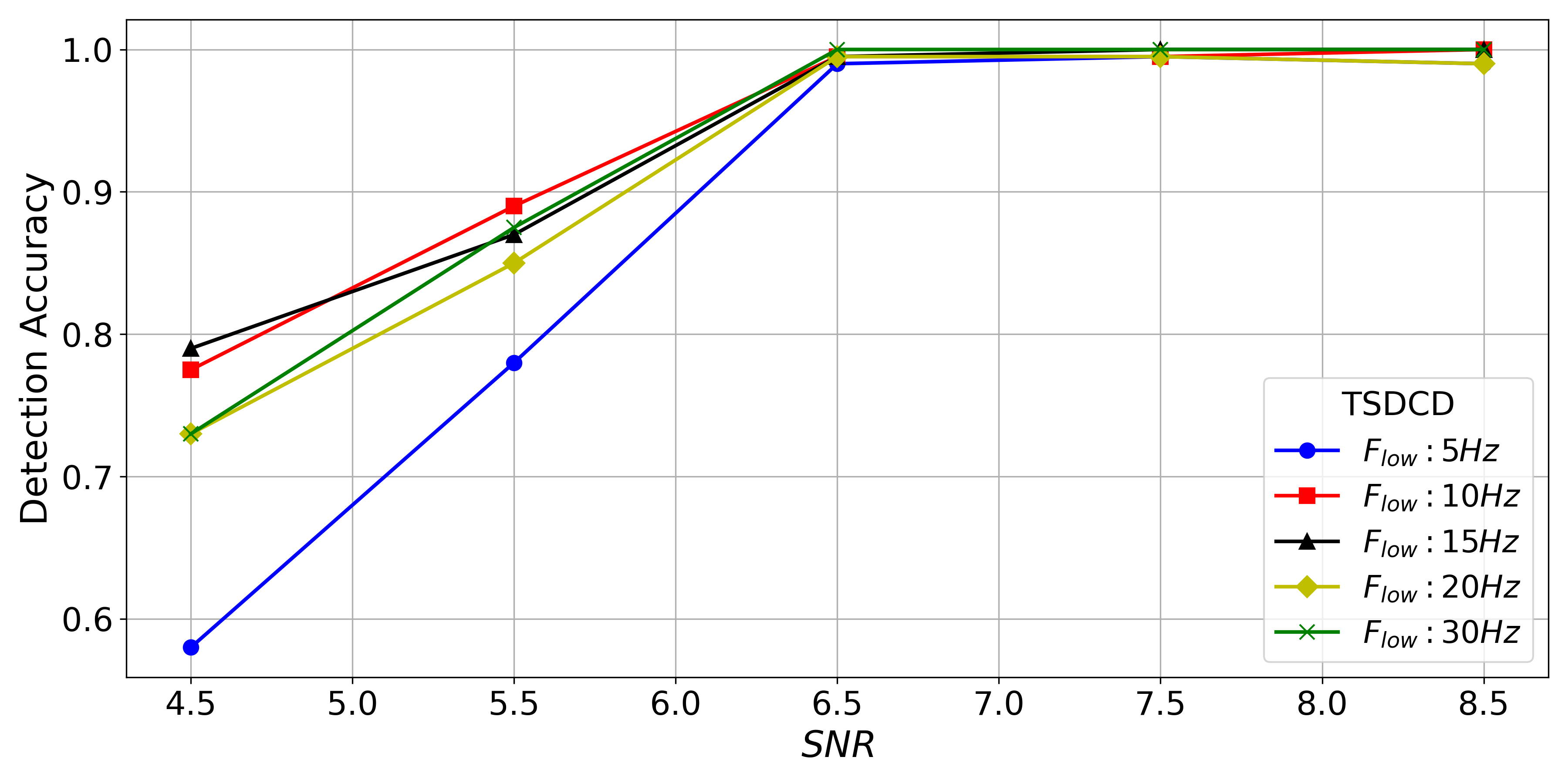}

  \caption{$MSNR$ versus detection's accuracy, Top: SSDD (from WTM1) versus TSDCD for sources with $F_{low}$ of 30 Hz. Bottom: TSDCD for sources with $F_{low}$ of 5 Hz, 10 Hz, 15 Hz, 20 Hz and 30 Hz.}
  \label{fig:acc}
\end{figure*}

\begin{figure*}[htb]
  \centering

\begin{minipage}{0.35\textwidth}
  \centering
  \includegraphics[width=\linewidth]{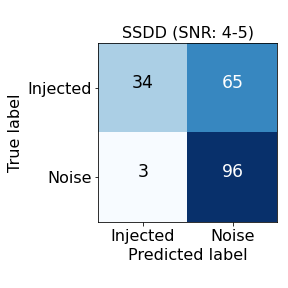}
\end{minipage}\hspace{1em}%
\begin{minipage}{0.31\textwidth}
  \centering
  \includegraphics[width=\linewidth]{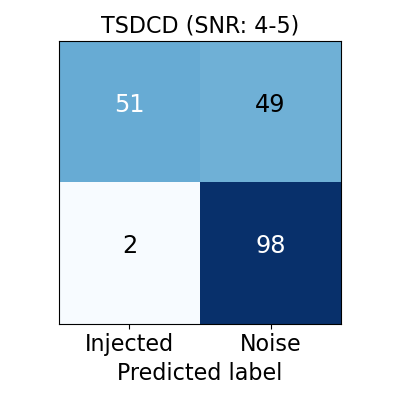}
\end{minipage}

\begin{minipage}{0.35\textwidth}
  \centering
  \includegraphics[width=\linewidth]{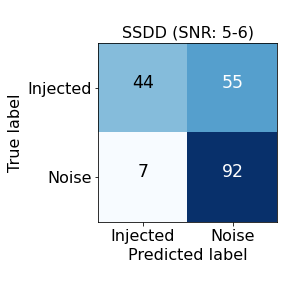}
\end{minipage}\hspace{1em}%
\begin{minipage}{0.31\textwidth}
  \centering
  \includegraphics[width=\linewidth]{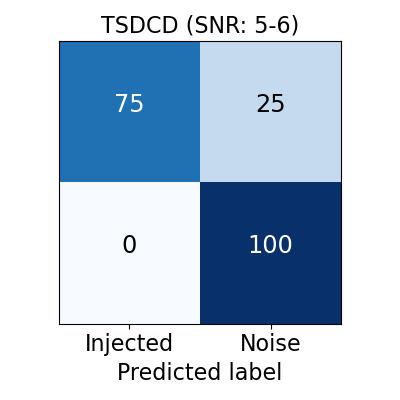}
\end{minipage}

  \begin{minipage}{0.35\textwidth}
    \centering
    \includegraphics[width=\linewidth]{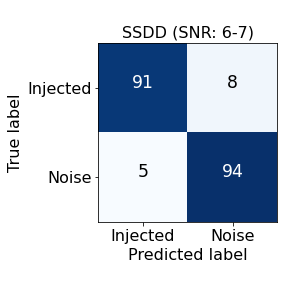}
  \end{minipage}\hspace{1em}%
  \begin{minipage}{0.31\textwidth}
    \centering
    \includegraphics[width=\linewidth]{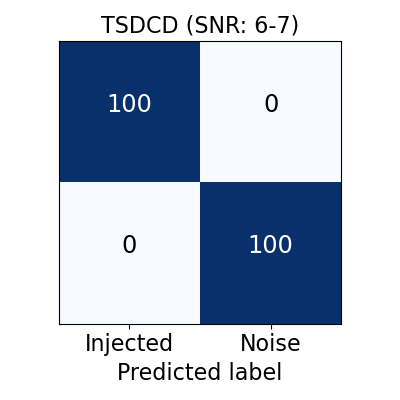}
  \end{minipage}

  \caption{Confusion matrix comparison for $MSNR$ in the ranges 4-7 between SSDD and TSDCD, with $F_{low}$ of 30 Hz.}
  \label{fig:cm1}
\end{figure*}
\begin{figure*}[h!]
    \centering
      \begin{minipage}{0.35\textwidth}
    \centering
    \includegraphics[width=\linewidth]{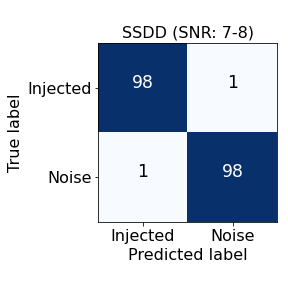}
  \end{minipage}\hspace{1em}%
  \begin{minipage}{0.31\textwidth}
    \centering
    \includegraphics[width=\linewidth]{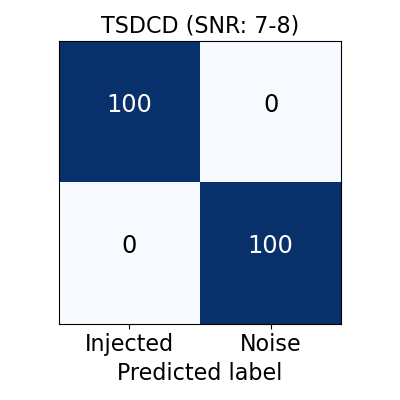}
  \end{minipage}
  \centering
    \begin{minipage}{0.35\textwidth}
    \centering
    \includegraphics[width=\linewidth]{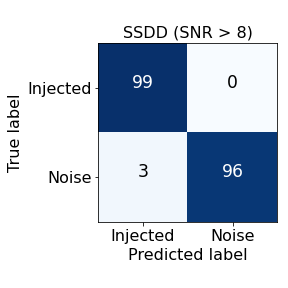}
  \end{minipage}\hspace{1em}%
  \begin{minipage}{0.31\textwidth}
    \centering
    \includegraphics[width=\linewidth]{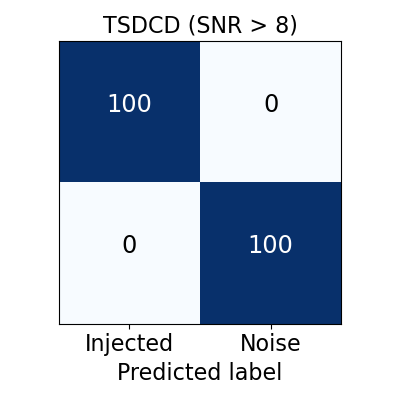}
  \end{minipage}

  \caption{Confusion matrix comparison for $MSNR$ 7-8 and greater between SSDD and TSDCD, with $F_{low}$ of 30 Hz.}
  \label{fig:cm2}
\end{figure*}

To evaluate the performance of ResNet on the TSDCD, similar to WTM1, we generated a classification report displaying precision, recall, and F1-score values \citep{10.5555/1162264} for our testing dataset, as shown in Table \ref{tab:report}. Recall, also known as sensitivity \citep{wang2020ecopann}, represents the quantity of correctly classified and misclassified sources, where ResNet achieved $80\%$ and $99\%$ on injected and only-noise spectrograms. This result demonstrates a notable trend, indicating that only-noise spectrograms are highly unlikely to be misclassified as injected ones across all the different $F_{low}$ and $MSNR$ ranges. Precision, assessing the quality of classification, achieved an average of $90\%$ by ResNet. The overall performance of ResNet, measured by the F1-score — a harmonic mean of precision and recall — also attained an average of $90\%$.


\begin{table}
\centering
\begin{tabular}{ccccc}
\hline
  Type & precision & recall & f1-score & support\\
  \hline
  Injected & 0.993 & 0.818 & 0.897 & 20,000\\ 
  Only noise  & 0.845  & 0.994 & 0.914 & 20,000\\
  \hline
  avg/ total & 0.919 & 0.906 & 0.905 & 40,000\\
  \hline
\end{tabular}
\caption{Classification report showing the precision, recall and F1-score metrics on the testing dataset.}
\label{tab:report}
\end{table}

To compare the TSDCD results with those obtained in our prior investigations in WTM1, we generated a TSDCD with an $F_{low}$ of 30 Hz only with all 5 $MSNR$ ranges. This dataset consists of 4000 sources, created using identical parameters to the testing dataset in WTM1. In Figure \ref{fig:acc}, the top panel displays the accuracy of the detection acquired for the TSDCD, alongside the accuracy obtained by SSDD in WTM1. The plot clearly demonstrates a significant improvement across all five $MSNR$ ranges, particularly at lower $MSNR$ values. The detection accuracy has improved from $60\%$, $60.5\%$, $84.5\%$, $94.5\%$, and $98.5\%$ to $78.5\%$, $84\%$, $99.5\%$, $100\%$, and $100\%$ for sources with $MSNR$ of 4-5, 5-6, 6-7, 7-8, and greater than 8, respectively. This shows a significant improvement of approximately $30.83\%$, $39.00\%$, $17.75\%$, $5.81\%$ and $1.52\%$ for sources with $MSNR$ of 4-5, 5-6, 6-7, 7-8, and greater than 8, respectively. It is essential to note that achieving $100\%$ accuracy for sources with $MSNR$ greater than 8 means successful detection of all sources from our samples. However, it does not guarantee that all sources within this specific range will be detected at all times. This applies to all other $MSNR$ ranges.

The bottom panel in Figure \ref{fig:acc} displays the detection accuracy from our testing dataset, considering each $F_{low}$ individually. The results indicate a generally similar performance across all frequencies and $MSNR$ ranges, with relatively inferior performance observed for lower $MSNR$ values (4-6) at $F_{low}$ of 5 Hz, as expected.





In a quantitative evaluation, we compared the number of misclassified injected and only-noise spectrograms using confusion matrices \citep{1997RSEnv..62...77S}. We used 200 randomly selected samples (100 samples from each class) from each $MSNR$ range, with an $F_{low}$ of 30 Hz, obtained from both SSDD and TSDCD to generate the confusion matrices. As shown if Figure \ref{fig:cm1}, a significant decrease can be observed in the number of misclassified Injected spectrograms from 66, 54, 8, and 1 to 50, 25, 0, and 0 for $MSNR$ 4-5, 5-6, 6-7, and 7-8, respectively, on the TSDCD, in contrast to the results from SSDD. For only-noise spectrograms, in TSDCD, only tow samples were misclassified as injected, in contrast to the 19 total misclassifications observed in SSDD. In Figure \ref{fig:cm2}, the confusion matrices for sources with $MSNR$ greater than 8 illustrate cases where no only-noise spectrogram was identified as injected in TSDCD, in contrast to 3 in SSDD.

\section{Evaluation on ET's continuous data}
\label{sec:eval2}

As previously done in WTM1, three sub-detectors combined mock data were generated to assess the efficiency of utilizing the TSDCD for near-real-time detection and to compare it with previously obtained results from SSDD. Five hours and 50 seconds of time-series data, with no overlapping sources and only BBH injections, encompassing all $MSNR$ ranges, were generated for each $F_{low}$ (5 Hz, 10 Hz, 15 Hz, 20 Hz, and 30 Hz) for each detector. At every five-second interval in this time-series data, a signal is injected at a random position, resulting in a total of 3,610 injected sources, with each $MSNR$ range containing 722 sources. The total duration of the data for each detector is 25 hours, 4 minutes, and 16 seconds. To compare with our previously obtained results, we utilize the data with an $F_{low}$ of 30 Hz.

\begin{figure*}[!ht]
    \centering

    \includegraphics[width=0.7\textwidth, height=0.3\textheight]{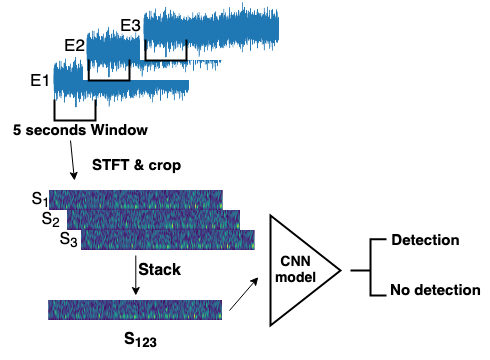}
    \caption{Inferencing on ET's continuous data}
    \label{fig:pipe}
\end{figure*}

To perform detection on this time-series data, we follow four main steps, as illustrated in Figure \ref{fig:pipe}: 1) slide a 5-second window simultaneously across the datasets of the three detectors, 2) generate a spectrogram for each window, 3) crop and stack spectrograms, and 4) feed the stacked spectrogram into the trained ResNet model for prediction.

The False Positive Rate (FPR) \citep{doi:10.1056/NEJM198810133191501} quantifies how often the model incorrectly classifies only-noise samples as injected ones, defined as follows:

\begin{equation}
    FPR = \frac{FP}{FP + TN}
\end{equation}

Where FP is the number of false positives, and TN is the number of true negatives.

\begin{figure*}[h]
  \centering

    \includegraphics[width=0.85\textwidth, height=0.3\textheight]{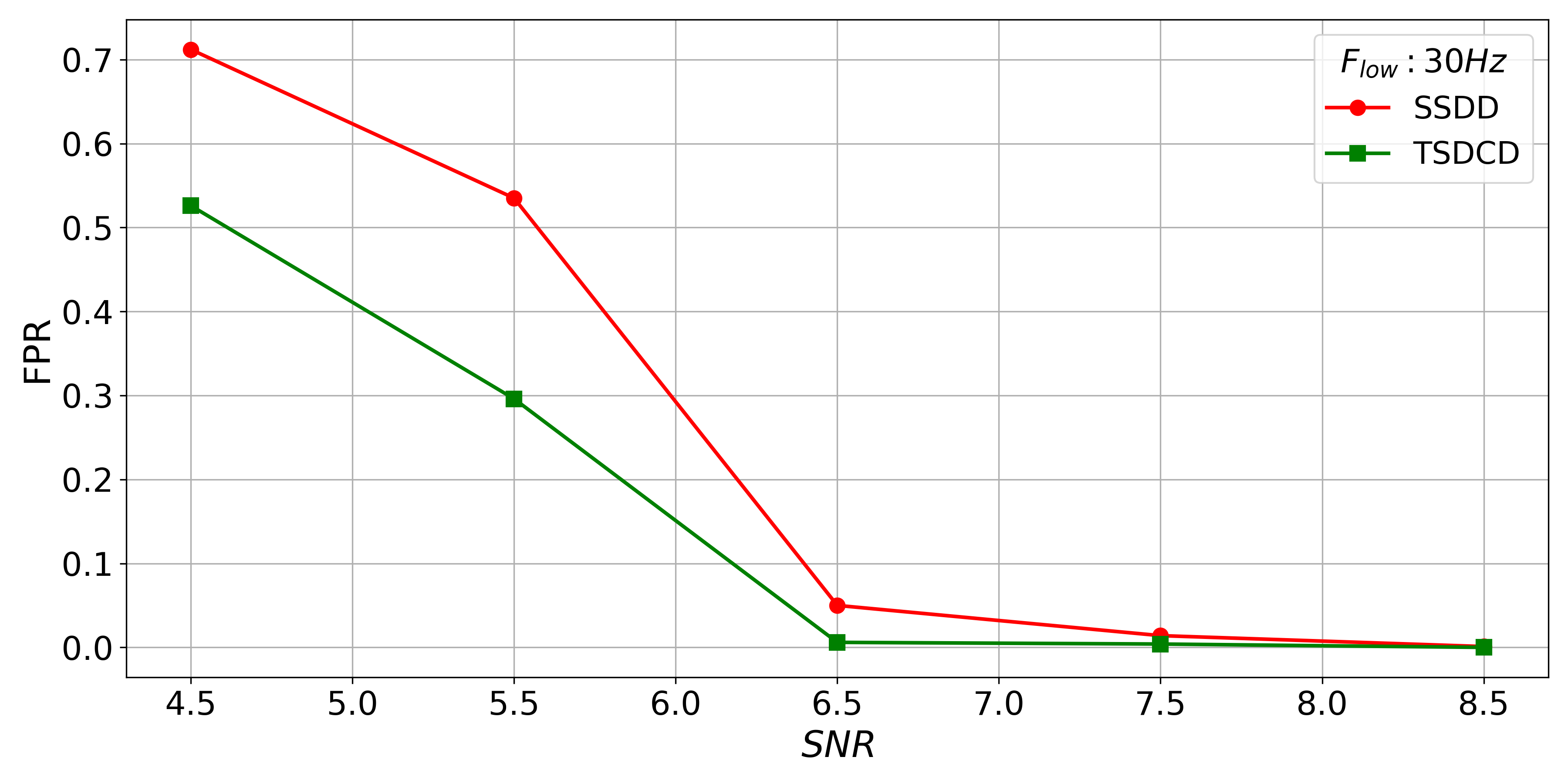}\\
    \includegraphics[width=0.85\textwidth, height=0.3\textheight]{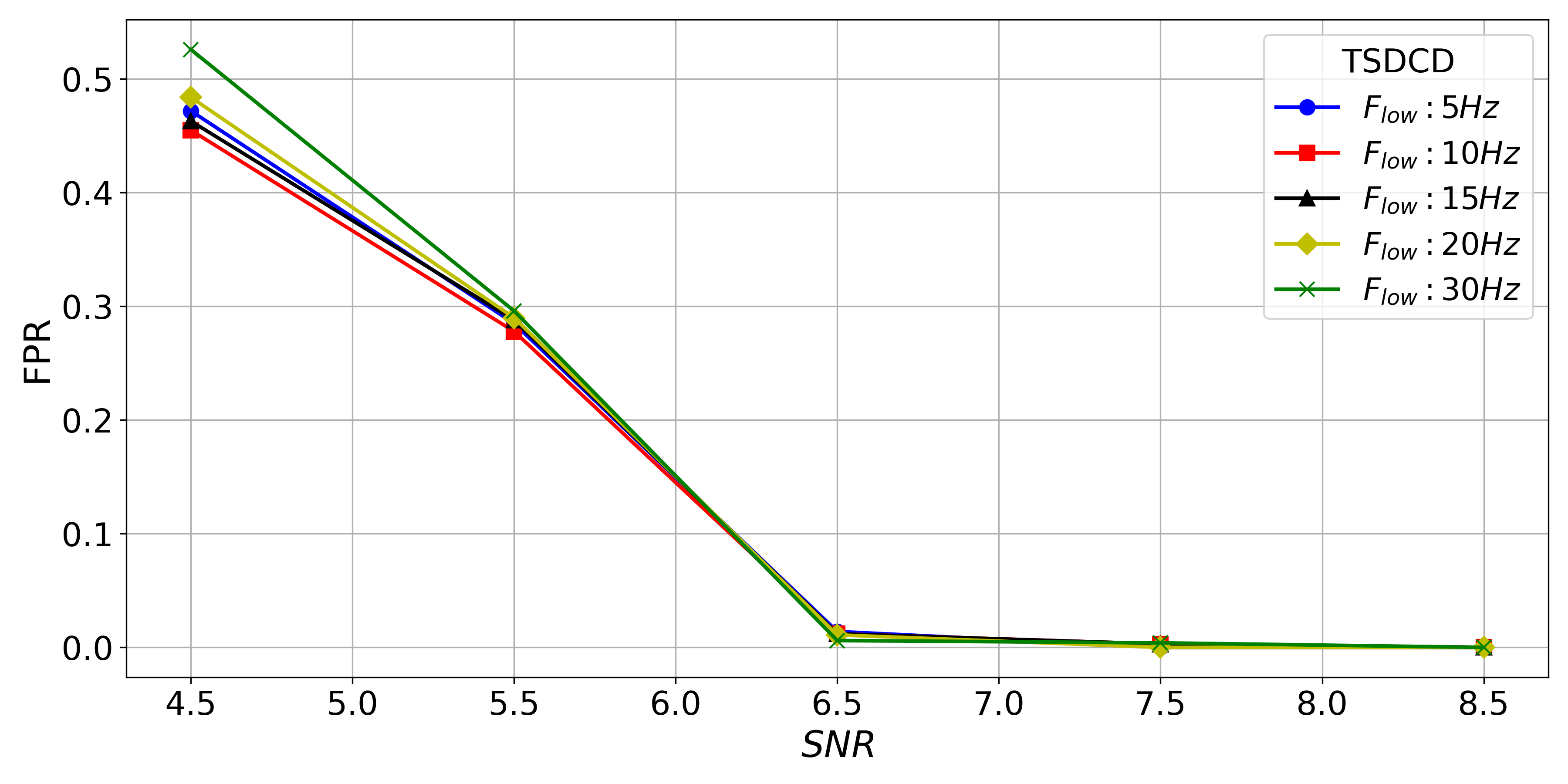}

  \caption{Top: A comparison of FPR values between SSDD (from WTM1) and TSDCD for sources with $F_{low}$ set at 30 Hz. Bottom: FPR values exclusively from TSDCD for sources with $F_{low}$ at 5 Hz, 10 Hz, 15 Hz, 20 Hz, and 30 Hz.}
  \label{fig:far_1_3arms}
\end{figure*}

\begin{figure*}[h]
  \centering
    \begin{minipage}{0.47\textwidth}
    \centering
    \includegraphics[width=\linewidth]{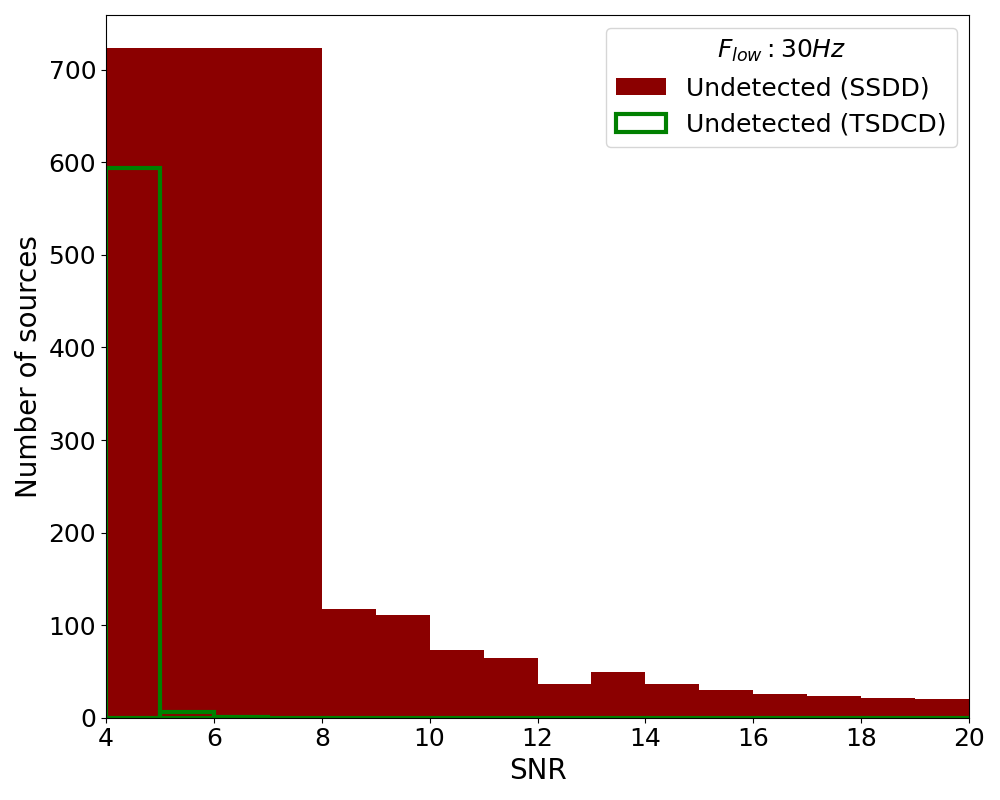}
  \end{minipage}\hspace{1em}%
  \begin{minipage}{0.47\textwidth}
    \centering
    \includegraphics[width=\linewidth]{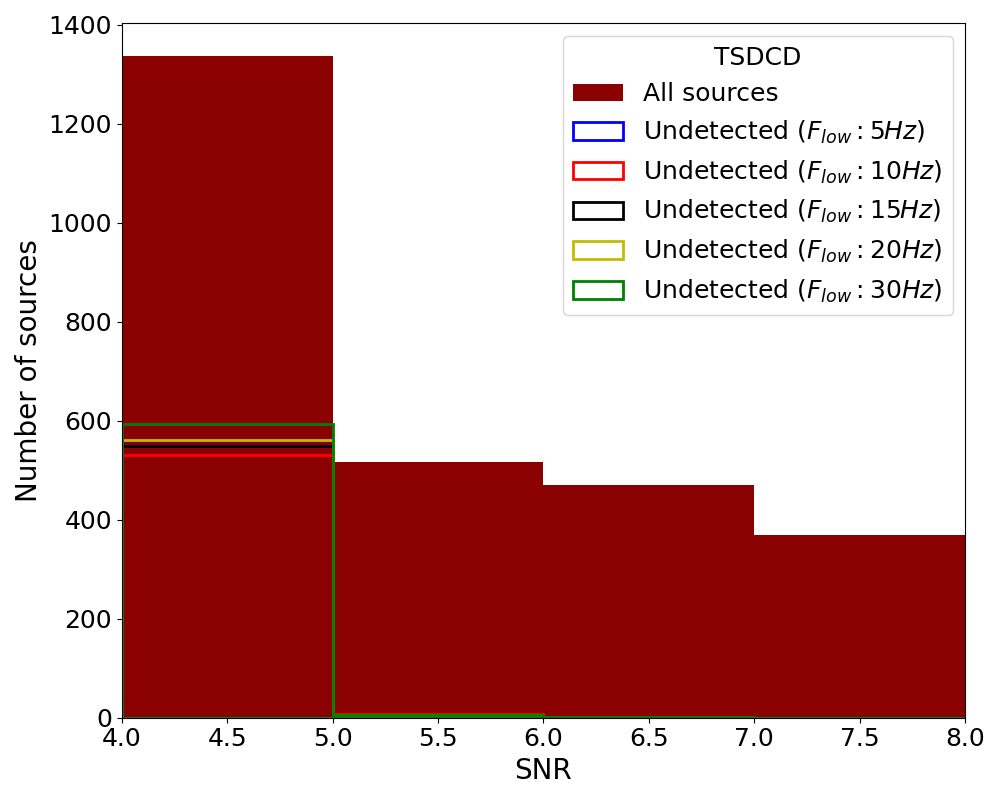}
  \end{minipage}

  \caption{Right: $MSNR$ of all undetected BBH sources from SSDD (from WTM1) and TSDCD. Left: $MSNR$ of undetected sources exclusively from TSDCD for sources with $F_{low}$ at 5 Hz, 10 Hz, 15 Hz, 20 Hz, and 30 Hz.}
  \label{fig:detection_snr}
\end{figure*}

\begin{figure*}[h]
  \centering
    \begin{minipage}{0.47\textwidth}
    \centering
    \includegraphics[width=\linewidth]{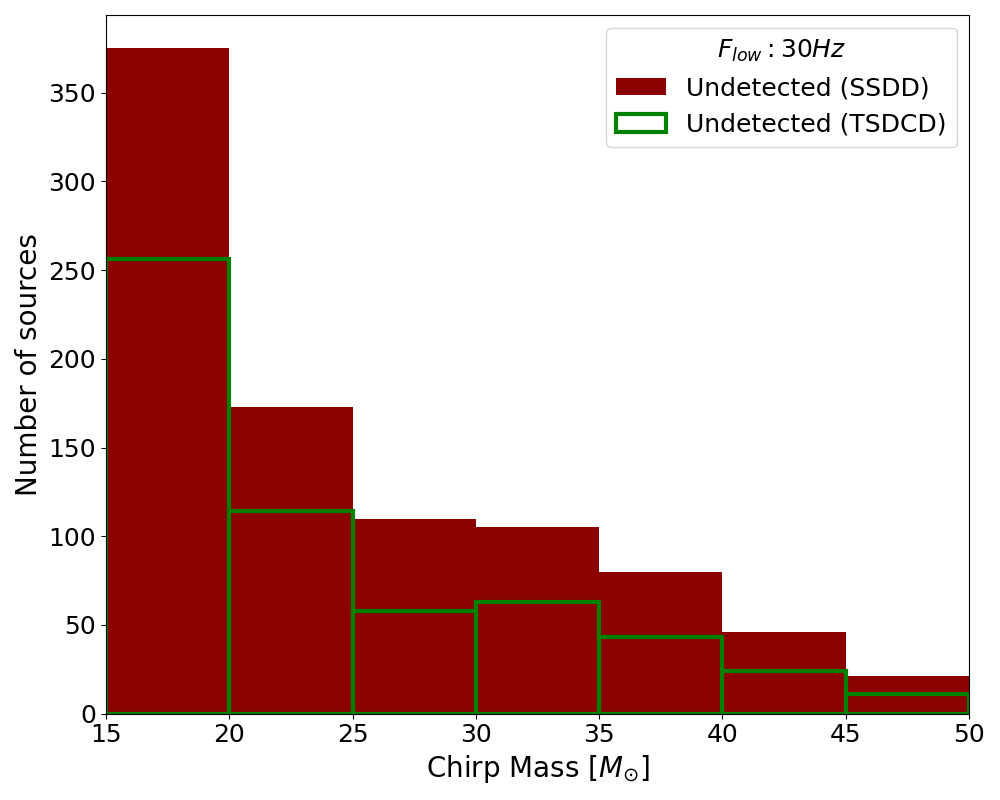}
  \end{minipage}\hspace{1em}%
  \begin{minipage}{0.47\textwidth}
    \centering
    \includegraphics[width=\linewidth]{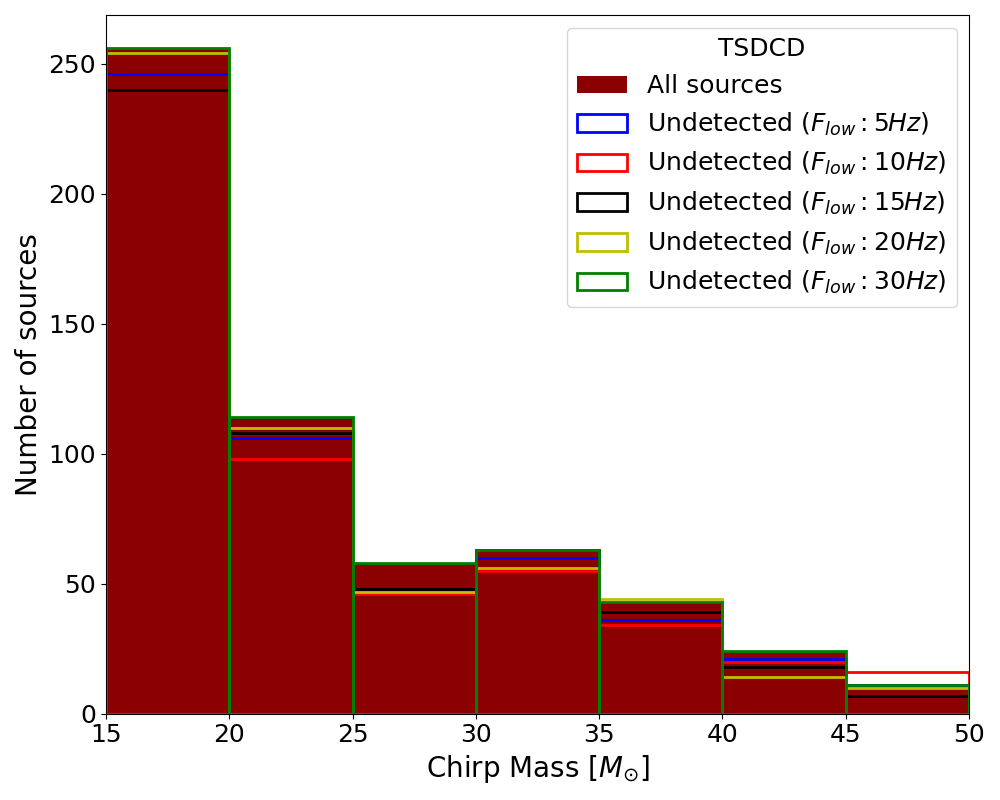}
  \end{minipage}

  \caption{Right: Chirp mass $M$ of all undetected BBH sources from SSDD (from WTM1) and TSDCD. Left: Chirp mass $M$ of undetected sources exclusively from TSDCD for sources with $F_{low}$ at 5 Hz, 10 Hz, 15 Hz, 20 Hz, and 30 Hz.}
  \label{fig:detection_m}
\end{figure*}

\begin{figure*}[tbp]
  \centering
    \begin{minipage}{0.47\textwidth}
    \centering
    \includegraphics[width=\linewidth]{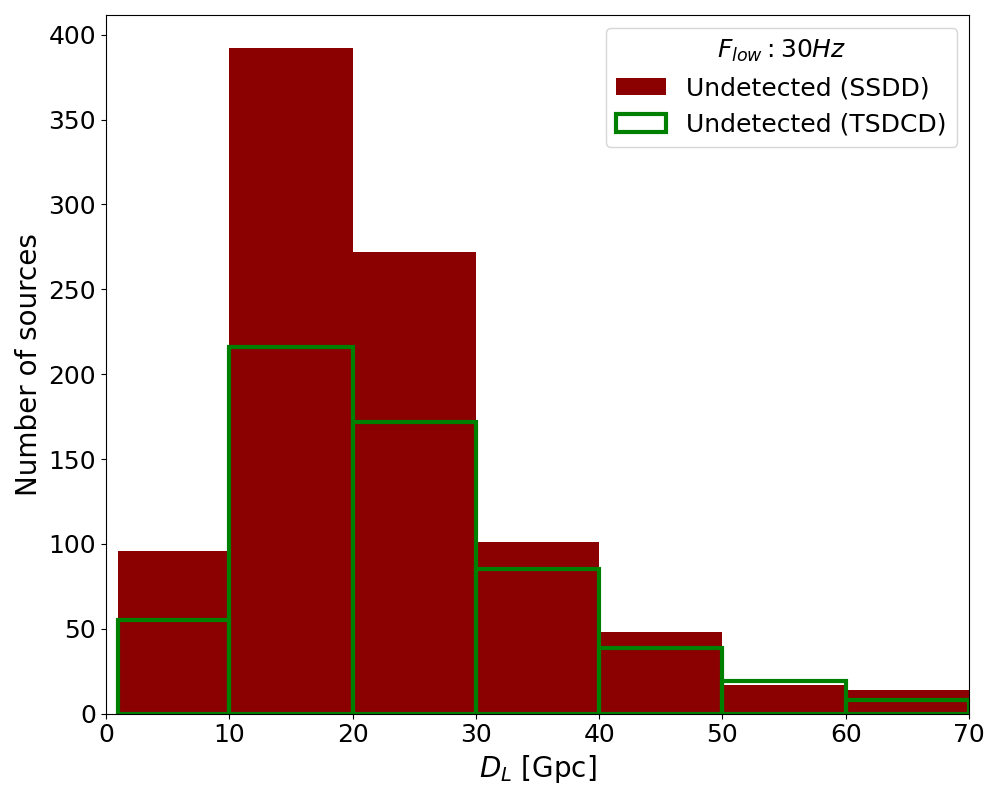}
  \end{minipage}\hspace{1em}%
  \begin{minipage}{0.47\textwidth}
    \centering
    \includegraphics[width=\linewidth]{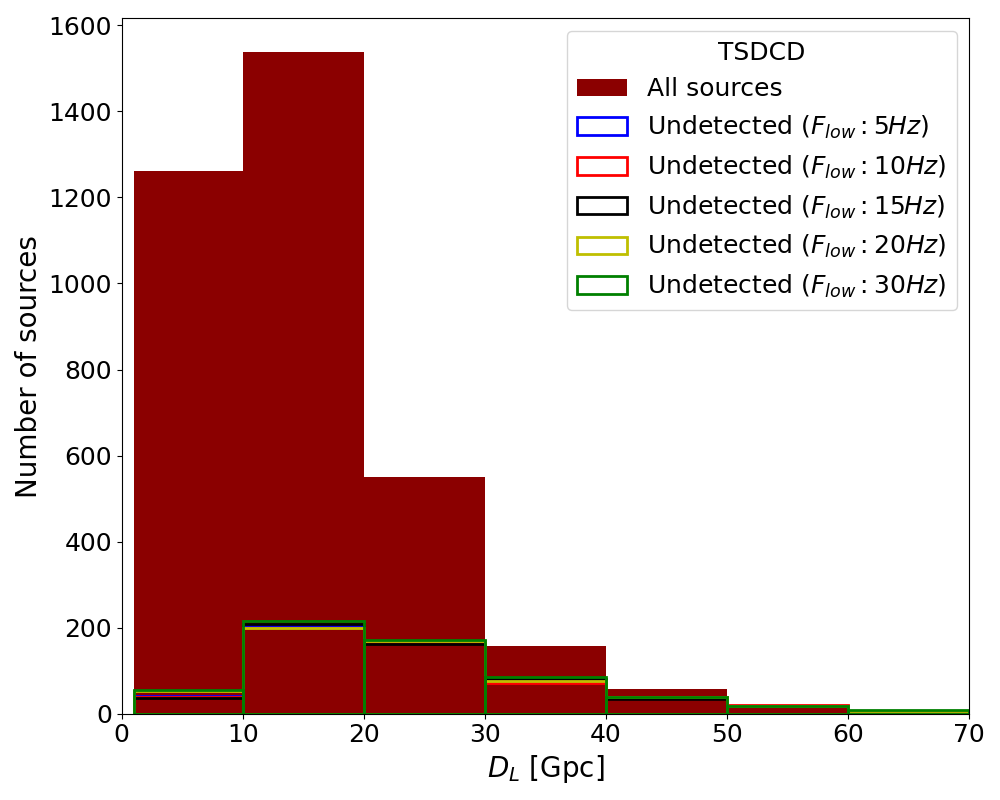}
  \end{minipage}

  \caption{Right: $D_{L}$ of all undetected BBH sources from SSDD (from WTM1) and TSDCD limited to 70 Gpc. Left: $D_{L}$  of undetected sources exclusively from TSDCD for sources with $F_{low}$ at 5 Hz, 10 Hz, 15 Hz, 20 Hz, and 30 Hz.}
  \label{fig:detection_dist}
\end{figure*}


FPR was computed for each $F_{low}$ for the entire mock data. Figure \ref{fig:far_1_3arms} depicts FPR as a function of $MSNR$, where in the top panel, TSDCD results are compared to SSDD results from WTM1. The general trend is consistent with the previous results in WTM1, where FPR decreases as $MSNR$ increases. As shown in the figure, FPR values have significantly improved, especially for sources with lower $MSNR$, transitioning from 0.712, 0.535, 0.050, 0.014, and 0.001 to 0.526, 0.296, 0.006, 0.004, and 0.0 for $MSNR$ ranges 4-5, 5-6, 6-7, 7-8, and >8, respectively. In the bottom panel, FPR values are depicted for each $F_{low}$. The overall performance is nearly identical across different $F_{low}$ settings, except for the case of $F_{low}$ at 5 Hz. This discrepancy aligns with the accuracy results presented earlier.

In terms of $MSNR$, chirp mass $M$, and luminosity distance $D_{L}$ for undetected sources, Figures \ref{fig:detection_snr}, \ref{fig:detection_m}, and \ref{fig:detection_dist} compare the total number of undetected sources between SSDD and TSDCD. Additionally, TSDCD is examined individually for all five different $F_{low}$ values. The right panel in Figure \ref{fig:detection_snr} demonstrates a significant improvement across all $MSNR$ ranges, particularly noticeable at relatively higher $MSNR$ 5-8 and greater, and noticeable decrease within $MSNR$ 4-5 where TSDCD exhibits a reduction of 100 undetected sources compared to SSDD. The left panel illustrates variations in the total number of undetected sources across our five $F_{low}$ settings, primarily within the $MSNR$ range of 4-5. Specifically, a total of 548, 531, 549, 562 and 594 undetected sources at $F_{low}$ settings of 5 Hz, 10 Hz, 15 Hz, 20 Hz, and 30 Hz is observed within $MSNR$ range 4-5, respectively. 

Similarly, in terms of chirp mass $M$, the right panel in Figure \ref{fig:detection_m} demonstrates a significant decrease in the total number of undetected sources in TSDCD, particularly among sources with chirp masses $M$ in the range of 15-20 \(M_\odot\). In TSDCD alone, count of undetected sources of different $F_{low}$ varies among all chirp mass $M$ values. For $F_{low}$ at 10 Hz, the minimum count observed is 240, 98, 46,55 and 34 in the chirp mass $M$ ranges of 15-20, 20-25, 25-30,30-35 and 35-40 \(M_\odot\), respectively. For the chirp mass $M$ ranges 40-45 and 45-50, the minimum count observed are 14 and 7 for $F_{low}$ of 20 Hz and 15 Hz, respectively.
Regarding the $D_{L}$, the right panel in Figure \ref{fig:detection_dist} shows a substantial reduction in the count of undetected sources in TSDCD, especially for sources with $D_{L}$ in the range of 10-30 Gpc. In the left panel, no significant variation is observed across different $F_{low}$ settings. In summary, both SSDD and TSDCD exhibit a higher detection rate for sources with higher $MSNR$, chirp mass, and shorter $D_{L}$, as indicated by smaller FPR values. Notably, TSDCD demonstrates a significant improvement over SSDD. It's worth noting that the same source may exhibit different $MSNR$ values on each sub-detector, yet TSDCD consistently outperforms SSDD across these variations. 

For a more qualitative check, Table \ref{tab:dud_report} shows the maximum $D_{L}$ ($D_{L_\text{Max}}$) in Gpc, minimum $MSNR$ ($MSNR_\text{Min}$), and the minimum chirp mass $M_\text{Min}$ in \(M_\odot\), along with the count of detected sources for all $F_{\text{low}}$ out of the total 3,610 sources. The ResNet model successfully detected sources at a distance of 86.601 Gpc, with an average $MSNR$ of 3.9 (averaged $MSNR$ over the three detectors) and a chirp mass of 13.632 at 5 Hz. At 10 Hz, 15 Hz, 20 Hz, and 30 Hz, ResNet successfully identified sources with average $MSNR$ values of 4.031, 4.033, 4.023, and 4.001, along with chirp masses of 20.320, 15.630, 14.201, and 17.532, respectively.
\begin{table}
\centering
\begin{tabular}{ccccc}
\hline
  $F_{\text{low}}$ & $D_{L_\text{Max}}$ & $MSNR_{Min}$ & $M_{Min}$ & Total\\
  \hline
  5 Hz & 86.601 & 3.900 & 13.632 & 3053 \\ \hline 

  10 Hz & 90.255 & 4.031 & 20.320 & 3069 \\ \hline 

 15 Hz & 67.553 & 4.033 & 15.630 & 3056 \\ \hline 

 20 Hz & 67.553 & 4.023 & 14.201 & 3042 \\ \hline 

 30 Hz & 67.553 & 4.001 & 17.532 & 3009 \\ \hline 
\end{tabular}
\caption{Maximum $D_L$ ($D_{L_\text{Max}}$) in Gpc, minimum $MSNR$ ($MSNR_\text{Min}$), minimum chirp mass ($M_\text{Min}$) in \(M_\odot\), and count of detected sources for all $F_{\text{low}}$ out of 3,610 total sources.}
\label{tab:dud_report}
\end{table}

In terms of computational efficiency, we concurrently processed a total of 25 hours, four minutes, and 16 seconds of data from each detector in parallel. Utilizing the same configuration as in WTM1 (Core i7 MacBook Pro with 16 GB memory, 2667 MHz DDR4, and 2.6 GHz processor), the entire data scan was completed in 15 minutes, averaging 1.9 minutes for each hour. The processing time has been significantly reduced compared with that achieved on SSDD of 4.7 minutes for each hour. The key difference lies in simultaneously reading data from all three sub-detectors. Hence, utilizing TSDCD is suitable for near-real-time detection and could be further enhanced with a more powerful setup.

\section{ET-MDC1: Einstein Telescope mock Data Challenge}
\label{sec:mdc}
A Mock Data Challenge for the Einstein Gravitational-Wave Telescope by \cite{2012PhRvD..86l2001R} was first released in 2012 and updated in early 2024 to contain 1.3 Terabytes of data for ET only, in addition to Cosmic Explorer. The recent release, called ET-MDC1, contains a continuous GW signal plus noise and noise only of about one month (30.8 days) split into 1300 segments of 2048 seconds, sampled at 8192 Hz. Using ET V-shaped detectors E1, E2, E3, GW signal and noise (colored Gaussian noise with no noise artifacts in this version) were simulated in addition to the null stream E0. The data contains the parameters of the injected sources, such as optimal SNR, component masses (M1 and M2) and $D_{L}$. The GW signal contains 59,540 BNS, 6,578 BBH, and 1,977 BHNS events, with optimal SNR ranges between 0.13 and 586.12, and For BNS, BBH and BHNS, IMRPhenomPNRv2 (which contains tidal effects), IMRPhenomXPHM and IMRPhenomXPHM approximant were used respectively. In ET-MDC1, BBH has a wide range of optimal SNR, spanning from 0.8 to 586. The component masses (M1 and M2) vary significantly, with minimum values of 7 and 6 \(M_\odot\) respectively, and maximum values reaching 793 and 617 \(M_\odot\). The $D_{L}$ of these BBH systems ranges from 0.5 Gpc to 154.37 Gpc.

The great challenge for our ResNet model to be evaluated on ET-MDC1 is that it was only trained on BBH mergers and with no overlapping sources. In addition to that, the waveforms were not continuous in the training dataset, and BNS has a long inspiral phase that can last for days, which the model has not seen during training. We evaluated ResNet on the entire ET-MDC1 E1, E2, and E3 injected data (signal plus noise) to assess its performance. Additionally, we tested the model on one week of noise-only data and null stream data to check for FPR with three different thresholds of 0.3 and 0.1 (equivalent to $50\%$, $70\%$ and $90\%$ confidence, respectively). As illustrated in Table \ref{tab:threshold_null_noise}, When accepting detection with at least $90\%$ confidence no FP was recorded. Only one false detection was observed using 0.3 threshold and 10 when using 0.5. 

\begin{figure*}[h]
  \centering
    \begin{minipage}{0.47\textwidth}
    \centering
    \includegraphics[width=\linewidth]{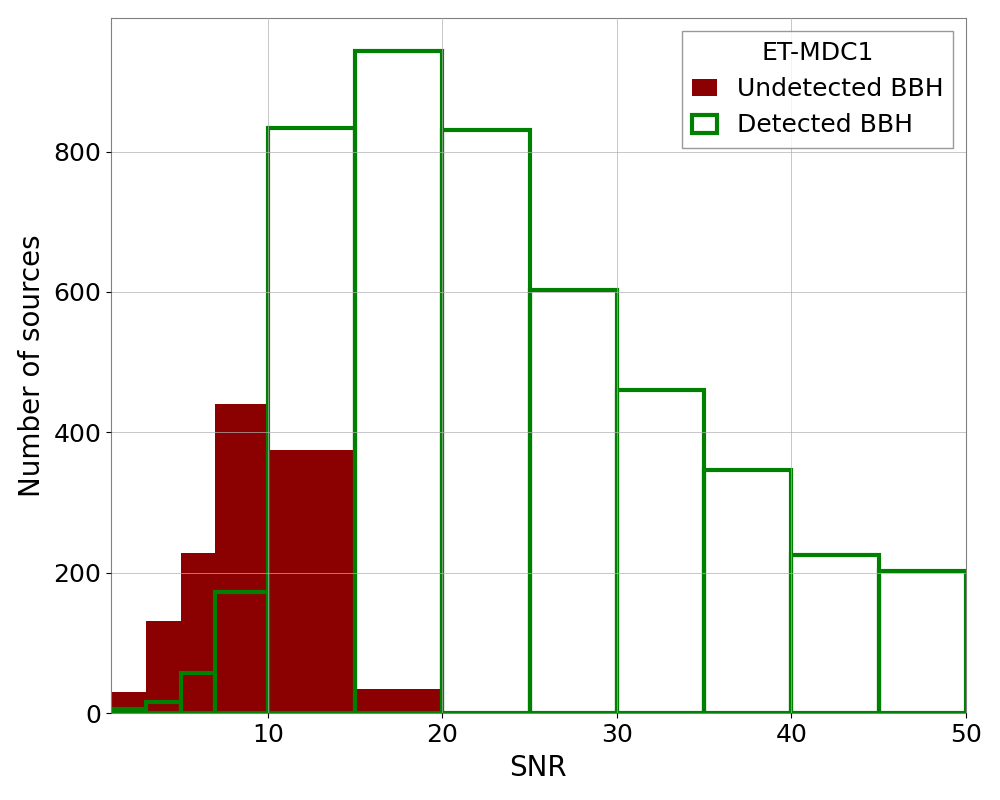}
  \end{minipage}\hspace{1em}%
  \begin{minipage}{0.47\textwidth}
    \centering
    \includegraphics[width=\linewidth]{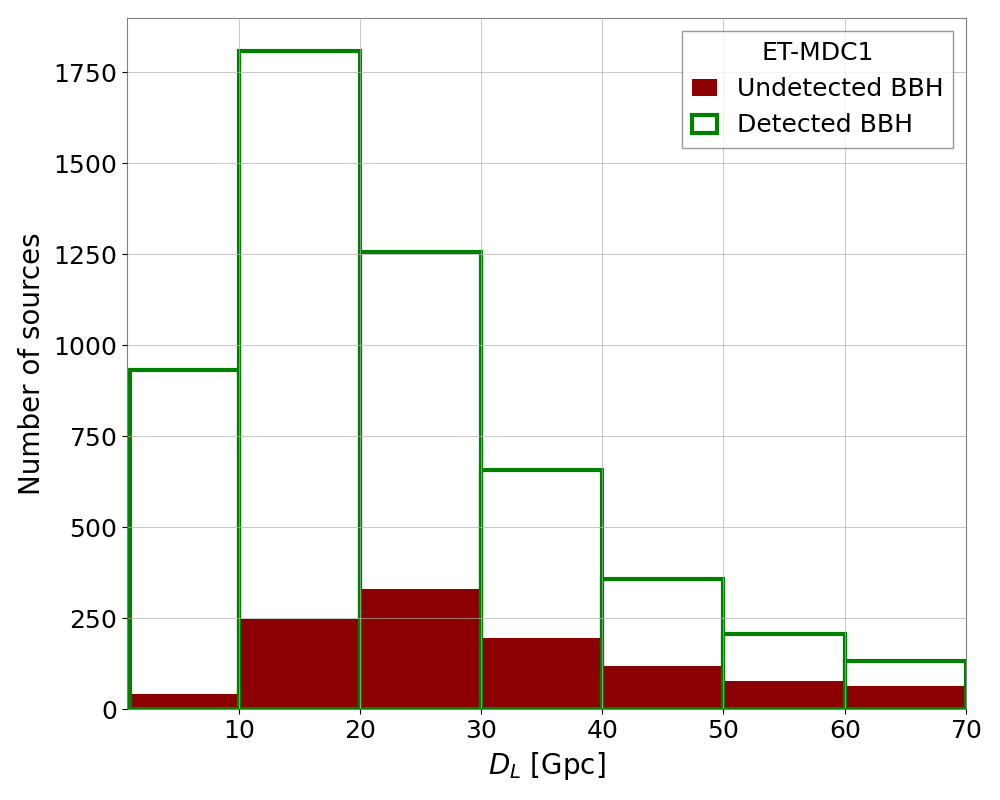}
  \end{minipage}

  \caption{Optimal SNR (left) and $D_{L}$ (right) of detected and undetected BBH sources from ET-MDC1.}
  \label{fig:mdc_dist}
\end{figure*}

\begin{figure*}[h]
  \centering
    \begin{minipage}{0.47\textwidth}
    \centering
    \includegraphics[width=\linewidth]{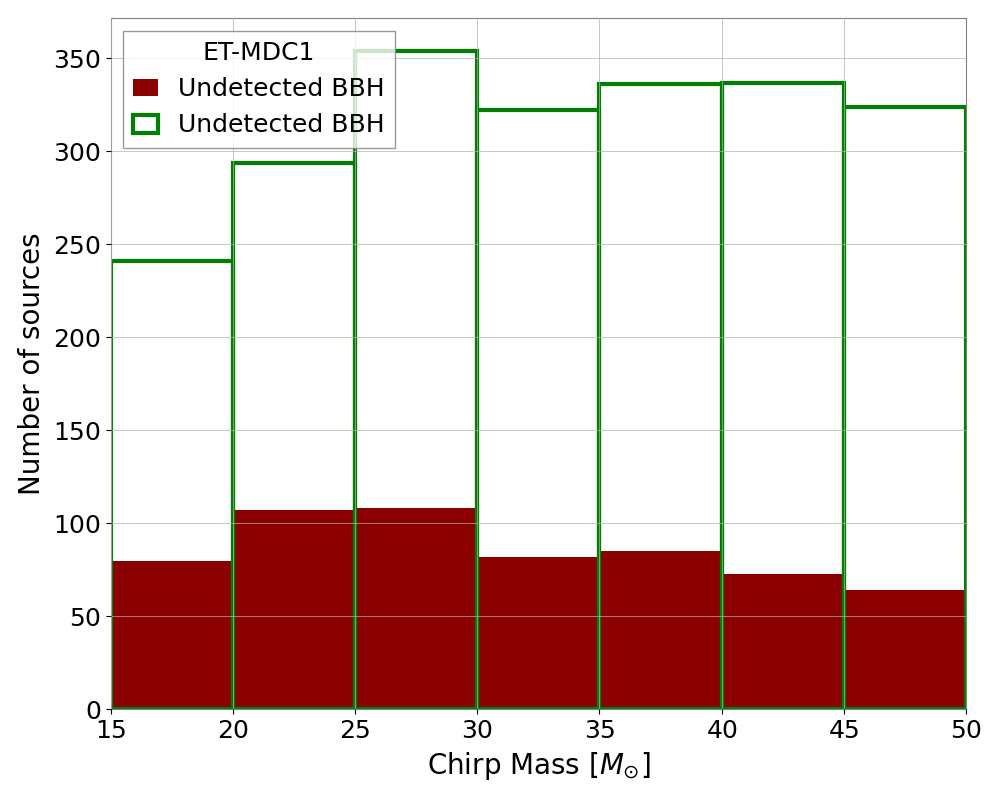}
  \end{minipage}\hspace{1em}%
  \begin{minipage}{0.47\textwidth}
    \centering
  \end{minipage}

  \caption{Chirp mass $M$ of detected and undetected BBH sources from ET-MDC1.}
  \label{fig:mdc_m}
\end{figure*}

\begin{table}[h!]
\centering
\begin{tabular}{|c|c|c|}
\hline
\textbf{Threshold} & \textbf{Null} & \textbf{Noise} \\
\hline
0.5 & 0 & 10 \\
\hline
 0.3 & 0 & 1 \\

\hline
0.1 & 0 &  0 \\
        
\hline
\end{tabular}
\caption{FPR from one week of noise and null data only.}
\label{tab:threshold_null_noise}
\end{table}

As shown in Figure \ref{fig:mdc_dist} and Figure \ref{fig:mdc_m}, ResNet successfully detected 5,566 BBH mergers out of a total of 6,578. Seventy-five percent of these sources have an average optimal SNR of 38.3, an average $D_{L}$ of 32 Gpc, and an average chirp mass of 96 \(M_\odot\). The detected BBH sources have a minimum and maximum $D_{L}$ of 0.5 Gpc and 148.95 Gpc, respectively. The minimum and maximum optimal SNR are 1.2 and 586, respectively, and the minimum and maximum chirp mass are 6 and 596 \(M_\odot\), respectively.

For undetected BBH sources, the optimal SNR ranges from 0.8 to 51.7, with seventy-five percent having an average optimal SNR of 10.7. The \(D_L\) ranges between 2.4 Gpc and 154.4 Gpc. The minimum and maximum chirp mass of the undetected sources are 7.8 \(M_\odot\) and 484.6 \(M_\odot\), respectively.

These results demonstrate the great performance of our model. The performance can be significantly improved by considering parameters outside the current range for BBH masses and distances. Additionally, incorporating samples of overlapping sources into the training dataset will positively impact the results.

\begin{figure*}[h]
  \centering
    \begin{minipage}{0.47\textwidth}
    \centering
    \includegraphics[width=\linewidth]{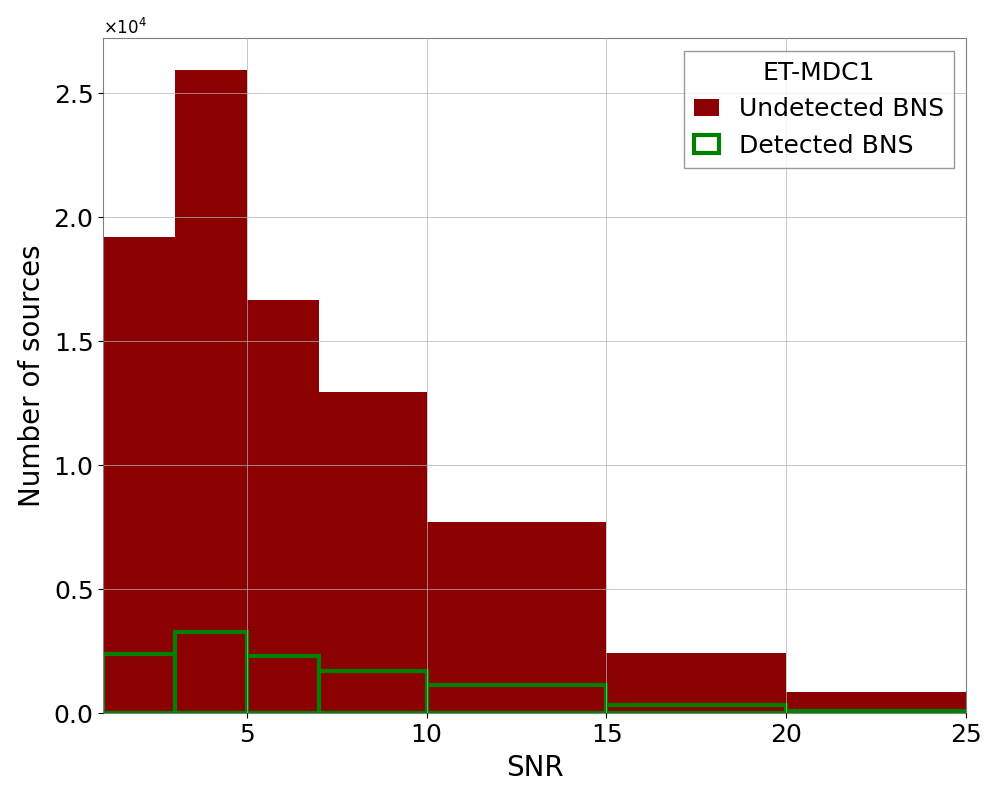}
  \end{minipage}\hspace{1em}%
  \begin{minipage}{0.47\textwidth}
    \centering
    \includegraphics[width=\linewidth]{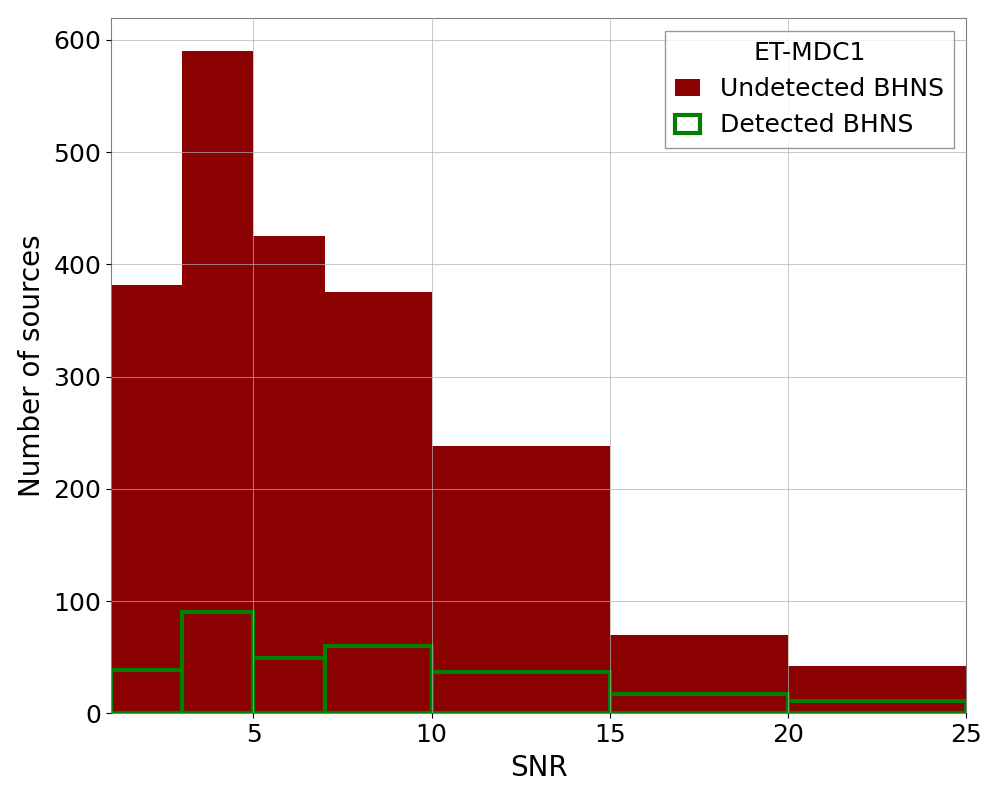}
  \end{minipage}

  \caption{Optimal SNR of detected and undetected BNS (right) and BHNS (left) sources from ET-MDC1.}
  \label{fig:mdc_snr_bhns}
\end{figure*}
Although ResNet was not trained on BNS and BHNS mergers, the model was able to detect 11,477 BNS mergers and 323 BHNS mergers. The optimal SNR of the detected and undetected BNS and BHNS sources are shown in the right and left panel of Figure \ref{fig:mdc_snr_bhns}. BNS and BHNS detected sources have optimal SNR ranges from 0.2 to 383 and from 1 to 50, respectively. 

The observation that the ResNet model is capable of detecting BNS and BHNS mergers, despite being trained only on BBH mergers, is due to the fact that inspiral GW signals share a common functional structure consisting of three phases: inspiral, merger, and ringdown. However, the exact form of the signal depends on several physical parameters such as the masses of the binary objects, their spins, the orbital eccentricity, and the distance to the source. When using matched filtering, a template generated with a specific set of parameters may perform poorly for detecting a signal with a different set of parameters, leading to a reduction in cross-correlation between the template and the observed signal \citep{macleod2016fully}. In contrast to matched filtering, deep learning models have the capacity to generalise beyond their training data. This characteristic allows CNNs to capture shared features across different types of compact binary systems, including BBH, BNS, and BHNS. All of these systems produce signals that include inspiral, merger, and ringdown phases, making the generalisation  ideal candidates for CNN-based classification. When training our model, we specifically focused on the merger phase, which is common across all compact binary systems. Our model was trained on thousands of merger waveforms with component masses ranging from 10 to 56 solar masses. This training allowed the network to learn features that generalise well across systems, irrespective of the specific mass or spin configurations. However, the performance can be significantly improved if this model is trained on diverse and well representatives samples of all three different systems, in addition to overlapping and non-overlapping sources.

\section{PyMerger}
\label{sec:pymerger}
PyMerger \citep{pymerger_2024} is a Python tool for detecting BBH mergers from ET, built based on the trained ResNet model described above. PyMerger is our first step towards developing a full AI-based pipeline for detection and, soon, parameter estimation of BBH from ET. We developed and shared the first version to open-source the trained model and allow results reproducibility. 
\begin{figure*}[!ht]
    \centering

    \includegraphics[width=0.7\textwidth, height=0.3\textheight]{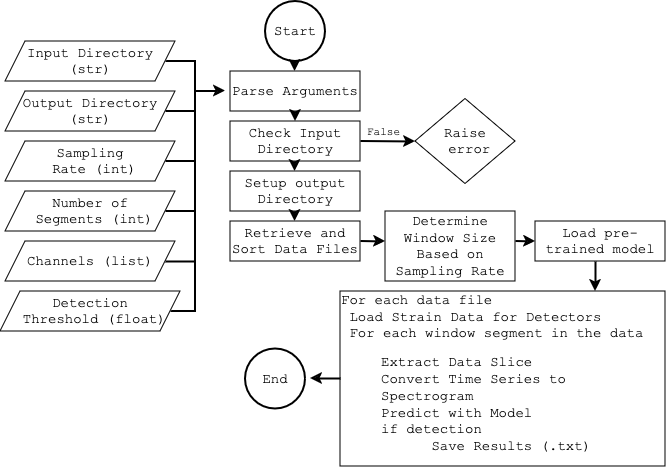}
    \caption{A flowchart outlines the high-level process of PyMerger, starting from argument parsing to data processing, model prediction, and result logging.}
    \label{fig:pymerger}
\end{figure*}

The current version handles only gravitational wave frame file format (.gwf), with the intention to support HDF5 in the upcoming version. As shown in Figure \ref{fig:pymerger}, the software takes seven arguments: 1) Directory containing the input .gwf files. 2) Directory to store the results. 3) Sampling-rate: Sampling rate of the input data (either 8192 or 4096). Default is 8192. 4) Number of data segments to be processed for each detector (i.e., number of .gwf files to be processed for each sub-detector). Default is 1. 5) List of the three channels names to be processed. Default is ['E1:STRAIN', 'E2:STRAIN', 'E3:STRAIN']. 6) Threshold value for merger detection. Default is 0.1 (accepting detection with at least $90\%$ confidence). The software processes each file by sliding a window (window size based on the sampling rate) to scan the data and look for BBH mergers. The output is a text file (.txt) with three columns: 1) Starting GPS time of the sliding window. 2) End GPS time of the sliding window. 3) Detection probability of all the detected mergers.  

Regarding computing efficiency, PyMerger scans one hour of TSDCD in 1.9 minutes. This can be significantly improved, as deep learning models can be substantially accelerated based on the hardware used. For instance, using the NVIDIA TensorRT-optimized AI ensemble, \cite{chaturvedi2022inference} processed one month of advanced LIGO data in 50 seconds. Future work will focus on incorporating BNS and BHNS events along with glitchy noise. 

The software is publicly available at \url{https://github.com/wathela/PyMerger}. Researchers in the field are encouraged to make use of the software and kindly provide us with their feedback, comments, and suggestions.
\section{Conclusion}
\label{sec:con}
In this study, we conducted a comparative analysis of the efficiency in detecting BBH sources using data from all three sub-detectors of ET simultaneously versus using data from a single sub-detector. The comparison was specifically focused on cases with $F_{\text{low}}$ set to 30 Hz. Furthermore, we explored TSDCD's performance across five different $F_{\text{low}}$ settings (5 Hz, 10 Hz, 15 Hz, 20 Hz, and 30 Hz), employing the same $MSNR$ ranges as in our previous study (4-5, 5-6, 6-7, 7-8, and >8). Utilizing the ResNet model, which exhibited superior performance in our previous study, TSDCD demonstrated a significant enhancement in detection accuracy compared to SSDD. The accuracy improved from $60\%$, $60.5\%$, $84.5\%$, $94.5\%$, and $98.5\%$ to $78.5\%$, $84\%$, $99.5\%$, $100\%$, and $100\%$ for sources with $MSNR$ ranges of 4-5, 5-6, 6-7, 7-8, and >8, respectively. The results indicate a substantial accuracy improvement for lower $MSNR$ ranges: 4-5, 5-6, and 6-7, with gains of $18.5\%$, $24.5\%$, and $13\%$ respectively. Additionally, there is a significant improvement of $5.5\%$ and $1.5\%$ for higher $MSNR$ ranges: 7-8 and >8. 

For more rigorous evaluation, ResNet model was evaluated on ET-MDC1 dataset, where the model demonstrated strong performance in detecting BBH mergers, identifying 5,566 out of 6,578 BBH events, with optimal SNR starting from 1.2, and a minimum and maximum $D_{L}$ of 0.5 Gpc and 148.95 Gpc, respectively. Despite being trained only on BBH mergers without overlapping sources, the model achieved high detection rates. However, to further enhance its performance, it is essential to include a broader range of parameters for BBH masses and distances, and incorporate overlapping sources in the training data. Notably, even though the model was not trained on BNS and BHNS mergers, it successfully detected 11,477 BNS and 323 BHNS mergers, with optimal SNR starting from 0.2 and 1, respectively. This indicates the model's potential for broader applicability. Future work will focus on improving detection rates by including both overlapping and non-overlapping BNS and BHNS events in the training process. Built on ResNet, PyMerger is a Python tool designed for detecting BBH mergers from ET data. PyMerger operates without the need for data whitening or band-passing and can scan one hour of TSDCD in 1.9 minutes on an average laptop. The current version supports GW frame format (.gwf) files.

%


\begin{acknowledgments}
This research was funded in part by National Science Centre, Poland \mbox{(UMO-2023/49/N/ST9/01287)}. Additionally, This work has been supported in part by the European Union’s Horizon 2020 research and innovation programme under grant agreement No 952480 (DarkWave project), and from the International Research Agenda Programme AstroCeNT \mbox{(MAB/2018/7)} funded by the Foundation for Polish Science (FNP) from the European Regional Development Fund. 
\end{acknowledgments}

%



\software{LALSuite \citep{lalsuite}, pyCBC \citep{alex_nitz_2022_5825666}}



\appendix

\section{ResNet: Deep Residual Neural Networks}
\label{sec:resnet}

Convolutional Neural Networks (CNNs) \cite{Fukushima1980} are deep learning models designed to process data with a 2D structure, like images or spectrograms, for tasks such as pattern recognition and object detection \cite{2015arXiv151108458O}. A CNN typically consists of convolutional layers that extract features using filters applied through convolution operations \cite{Goodfellow-et-al-2016}, followed by activation functions (e.g., ReLU, $tanh$, $softmax$) \cite{10.5555/521706, 10.1007/978-3-642-76153-9_28}. Pooling layers reduce the spatial dimensions by aggregating local regions, and fully connected layers act as classifiers, similar to traditional Artificial Neural Networks (ANNs) \cite{2017arXiv171203541A}. In essence, CNNs perform non-linear mappings from input images to class scores. CNNs model suffer from vanishing gradient problem (also called degradation problem), where increasing the depth of the network leads to higher training error. When deep neural networks are trained, an increase in network depth often leads to a saturation and subsequent degradation in performance. The degradation is not caused by overfitting but by difficulties in optimization. This is counterintuitive because deeper models are expected to, theoretically, learn better representations.

\begin{figure*}[!ht]
    \centering
    \includegraphics[width=0.6\textwidth, height=0.3\textheight]{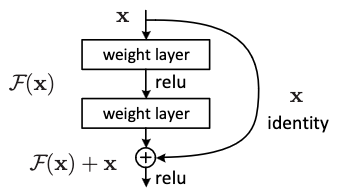}
    \caption{Residual learning: a residual block.}
    \label{fig:resnet}
\end{figure*}

Deep Residual Neural Networks (ResNets) were introduced by \cite{https://doi.org/10.48550/arxiv.1512.03385} as a solution to the degradation problem in deep learning models. ResNets solve this problem by introducing shortcut connections that bypass certain layers, allowing for easier optimization of deep models. ResNet introduces the concept of residual learning.The core component of ResNet is the \textit{residual block}, shown in Figure \ref{fig:resnet}, where the network learns a \textit{residual function} instead of directly learning the desired mapping. Let \( X \) represent the input to a residual block and \( \mathcal{F}(X) \) be the desired transformation. In a traditional network, the output of a layer would be \( H(X) = \mathcal{F}(X) \). However, in ResNet, the network learns a residual function \( \mathcal{R}(X) \) defined as:

\begin{equation}
\mathcal{R}(X) = \mathcal{F}(X) - X,
\end{equation}

which can be rearranged as:

\begin{equation}
\mathcal{F}(X) = \mathcal{R}(X) + X.
\end{equation}

This residual function \( \mathcal{R}(X) \) is learned by a series of layers, and the output of the residual block is:

\begin{equation}
Y = X + \mathcal{R}(X).
\end{equation}

The addition of the input \( X \) to the learned residual mapping allows the network to bypass transformations, helping to preserve gradients during backpropagation.

By stacking multiple residual blocks, ResNet enables the training of very deep networks without the performance degradation seen in traditional deep architectures. The general form for the output of the \( l \)-th residual block in a deep ResNet can be written as:

\begin{equation}
Y^{[l+1]} = Y^{[l]} + \mathcal{F}(Y^{[l]}, W^{[l]}),
\end{equation}

where \( Y^{[l]} \) is the input to the \( l \)-th block, \( \mathcal{F}(Y^{[l]}, W^{[l]}) \) is the residual function learned by the \( l \)-th block, and \( W^{[l]} \) are the trainable weights of the block. For deeper networks like ResNet-50 or ResNet-101, the bottleneck design is used to reduce the number of parameters. In this design, three layers are used instead of two. These layers use a combination of three convolutions: 1) A 1x1 convolution to reduce the dimensionality 2) A 3x3 convolution for feature extraction 3) A final 1x1 convolution to restore the original dimensionality. The residual function for a bottleneck block is:

\begin{equation}
\mathcal{R}(X) = W_3 \cdot (W_2 \cdot (W_1 \cdot X)),
\end{equation}

where \( W_1 \), \( W_2 \), and \( W_3 \) are the weights for the three convolutional layers. The bottleneck design reduces computational complexity by reducing the dimensionality of the feature maps before applying expensive convolutions. When the dimensions of the input \( X \) and the output \( \mathcal{F}(X) \) do not match, ResNet employs a \textit{projection shortcut} to match dimensions using a linear projection:

\begin{equation}
Y = W_s \cdot X + \mathcal{F}(X),
\end{equation}

where \( W_s \) is the projection matrix (typically a 1x1 convolution) used to adjust the dimensions of \( X \) before adding it to the residual function. For the training, ResNet uses standard backpropagation \citep{Munro2010}, with the total loss \( \mathcal{L} \) minimized using stochastic gradient descent (SGD). For classification tasks, the loss function is typically cross-entropy:

\begin{equation}
\mathcal{L} = - \sum_i y_i \log(\hat{y}_i),
\end{equation}

where \( y_i \) is the true label and \( \hat{y}_i \) is the predicted probability for class \( i \). The introduction of residual connections significantly improved the performance of deep neural networks. ResNet achieved state-of-the-art results in the ImageNet competition \citep{russakovsky2015imagenet}, and has been widely adopted in various applications, including image classification, object detection, and segmentation.

\bibliography{sample631}{}
\bibliographystyle{aasjournal}



\end{document}